\documentclass{rQUF2e}

\usepackage{epstopdf}
\usepackage{subfigure}

\usepackage{setspace} 

\usepackage{natbib}
\usepackage{dsfont}
\usepackage{bigints}
\usepackage{tikz}
\usetikzlibrary{intersections,decorations.markings}
\usetikzlibrary{shapes}
\usepackage{cases}
\usepackage{csquotes}
\bibpunct[, ]{(}{)}{,}{a}{}{,}%
\def\bibfont{\small}%
\def\bibsep{\smallskipamount}%
\def\bibhang{24pt}%
\usepackage{url}
\usepackage[]{algorithm2e,algpseudocode}
\usepackage{multirow}
\usepackage{listings}

\usepackage{mathtools}
\DeclarePairedDelimiter{\ceil}{\lceil}{\rceil}

\theoremstyle{plain}

\theoremstyle{definition}

\theoremstyle{remark}
\newtheorem{remark}{Remark}

\begin{document}


\title{An SFP--FCC Method for Pricing and Hedging Early-exercise Options under L\'evy Processes}

\author{Tat Lung (Ron) Chan$^{\ast}$$\dag$\thanks{$^\ast$Corresponding author.
Email: t.l.chan@uel.ac.uk} \\
\affil{$\dag$School of Business, University of East London, Water Lane, Stratford, UK, E15 4LZ\\} }

\maketitle

\begin{abstract}
This paper extends the Singular Fourier--Pad\'e (SFP) method proposed by \citet{Chan:2018} to pricing/hedging early-exercise options--Bermudan, American and discrete-monitored barrier options--under a L\'evy process. The current SFP method is incorporated with the Filon--Clenshaw--Curtis (FCC) rules invented by \citet{Dom_Gra:2011}, and we call the new method SFP--FCC. The main purpose of using the SFP--FCC method is to require a small number of terms to yield fast error convergence and to formulate option pricing and option Greek curves rather than individual prices/Greek values.  We also numerically show that the SFP--FCC method can retain a global spectral convergence rate in option pricing and hedging when the risk-free probability density function is piecewise smooth. Moreover, the computational complexity of the method is $\mathcal{O}((L-1)(N+1)(\tilde{N} \log \tilde{N}) )$ with $N$ a (small) number of complex Fourier series terms, $\tilde{N}$ a number of Chebyshev series terms and $L$, the number of early-exercise/monitoring dates. Finally, we show that our method is more favourable than existing techniques in numerical experiments.
\end{abstract}


\begin{classcode}C6, C63\end{classcode}

\section{Introduction}\label{sec:intro}

A Bermudan option can be exercised on predetermined dates before maturity. The option holder receives the exercise payoff when he/she exercises the option on specific dates at the option's maturity. Between two consecutive exercise dates, the valuation process can be regarded as similar to
a European option, which can be priced and hedged using the risk-neutral valuation formula \citep[cf.][]{Chan:2018, Cha_Hal:2019}.

If we consider $\log S_t:=x_t$ driven by a L\'evy process and a Bermudan option with strike $K$ and maturity $T$ that can be exercised only on a given number of exercise dates $t=t_0 <t_1\leq t_2\leq\ldots t_l\leq t_{l+1}\leq\ldots\leq t_L = T.$, we can write the risk-neutral Bermudan pricing formula for such an option as
\begin{align}\label{eqn:Bermuda_intro1}
V(x_{t_l},K,t_l)&=\begin{cases} U(e^{x_{t_l}},K,t_l) & l=L,\, t_L=T\\
\max\left(C(x_{t_l},K,t_l), U(e^{x_{t_l}},K,t_l) \right) & l=1,2,3,\dots,L-1\\
C(x_{t_l},K,t_l)&l=0
\end{cases},
\end{align}
where, $U(e^{x_{t_l}},K,t_l)$ is the payoff function at $t_l.$, i.e., if the payoff function is a call, then $U(e^{x_{t_l}},K,t_l)$ is transformed into $\max\left(e^{x_{t_l}}-K,0\right).$ In (\ref{eqn:Bermuda_intro1}), $C(x_{t_l},K,t_l)$ at each $t_j$ can be described as a risk-neutral valuation formula:
\begin{align}\label{eqn:Bermuda_intro2}
C(x_{t_j},K,t_j)&=e^{-r(t_{j+1}-t_j)}\mathbb{E}\left(V(x_{t_{j+1}},K,t_{j+1}) \vert x_{t_j} \right)\nonumber\\
&=e^{-r(t_{j+1}-t_j)}\int_{-\infty}^{+\infty} V(e^{x + \chi-\log K}, t_{j+1}) f(\chi) \mathrm{d} \chi,\quad \chi \in X_{t_{j+1}}-X_{t_j}.
\end{align}
Here, $X_{t_{j+1}}-X_{t_j}$ is the L\'evy process, $r$ is the risk-neutral interest rate, and $f(\chi)$ is the risk-neutral probability density function (PDF). As (\ref{eqn:Bermuda_intro2}) is an expectation and integral, a sustainable number of numerical methods are developed to calculate it. The popular methods include, for example, the FFT--QUAD method, a combination of the Fast Fourier Transform (FFT) method and numerical quadrature, suggested by \citet{Sul:2005}; the CONV method, an FFT method proposed by \cite{Lor_Fan:2008}; a mixture of the FFT method and the Guass transform \citep[e.g.][]{Bro_Yam:2003} or the Hilbert transform \citep[e.g.][]{Fen_Vad:2008,Zen_Kwok:2014}; the COS method, a Fourier-cosine series approach suggested by \cite{Fan_Oos:2009b}; and the SWIFT method, a wavelet series approach \citep{Mar:2015, Mar_Gra:2017}. The advantage of using the FFTs, COS and SWIFT methods for option pricing is that they can achieve a global spectral (exponential) convergence rate and require fewer summation terms as long as the governing PDF is sufficiently smooth. However, when the difference $\Delta t$ between $t_{j}$ and $t_{j+1}$ approaches zero in (\ref{eqn:Bermuda_intro2}), $f(\chi)$ tends to become highly peaked and piecewise continuous (non-smooth)\footnote{A function is called piecewise continuous on an interval if the function is made up of a finite number of $\nu$ times differentiable continuous pieces.} in any L\'evy process.  Using any type of Fourier series to represent a piecewise continuous function, e.g., a piecewise continuous PDF, is notoriously fraught and causes the Gibbs phenomenon \citep[cf.][]{Dris_Ben:2001, Dris_Ben:2011}. The impact of the Gibbs phenomenon can lead to inaccurate pricing and hedging and a lack of spectral convergence when the approximate option prices are generated via FFT or Fourier series methods at or around the jumps.

Accordingly, we propose the singular Fourier-Pad\'e (SFP) method \citep{Chan:2018} to circumvent the mentioned problem to allow fewer summation terms and maintain spectral convergence when $f(\chi)$ is piecewise continuous.
Why do we choose the SFP method? We exhibit the following characteristics when we use the method to price and hedge European-type options:
\begin{enumerate}
\item a global spectral convergence rate for piecewise continuous PDFs;
\item fast error convergence with fewer partial summation terms required;
\item accurate pricing of any European-type option with the features of deep in/out of the money and very long/short maturities;
\item consistent accuracy for approximating large or small option prices throughout.
\end{enumerate}

To obtain the same advantages of using the SFP method, we extend the current method with the help of the Filon--Clenshaw--Curtis (FCC) rules, invented by \citet{Dom_Gra:2011}, to price Bermuda options and American and discrete-monitored barrier options. We call the new method SFP--FCC. Compared with the SFP method alone, the main advantage of the SFP--FCC method is that it can not only require fewer summation terms to yield spectral convergence with a (piecewise) continuous PDF but also provide option pricing and an optional Greek formula rather than individual prices/Greek values.

The remainder of this paper is structured as follows. Section~\ref{sec:intro} provides an introduction. Section~\ref{sec:SFP} describes the SFP method. Section~\ref{sec:Fin_Model} introduces the financial stochastic models that we examine in this paper. Section~\ref{sec:OptionFormu} revises and improves the formulation of the SFP option pricing formulae for European options proposed in \citet{Chan:2018}. In Section~\ref{sec:EEOptions}, we propose the SFP--FCC algorithms/formulae to price Bermudan, American (cf. Section~\ref{sec:Bermudan}) and discrete-monitored barrier options (cf. Section~\ref{sec:barrier}) and to find an early-exercise point by using root-finding techniques (cf. Section~\ref{sec:Newton}). Section~\ref{sec:greek_truc} describes the derivation of the option Greek formulae and the choice of truncated integration intervals. Section~\ref{sec:results} discusses, analyses and compares the numerical results of the SFP--FCC method with the results of other numerical methods. We conclude and discuss possible future developments in Section~\ref{sec:conclusion}. Finally, Appendix \ref{sec:SFP_algorithm} shows the algorithm of computation of the SFP coefficients, and Appendix \ref{sec:sing} discusses the method of locating jumps in PDFs. Appendix \ref{sec:weights} describes the FCC rules, and Appendix \ref{sec:cums} shows the table of cumulants.

\section{Singular Fourier--Pad\'e interpretation and correction of the Gibbs phenomenon}\label{sec:SFP}
If we consider a function $f$ with a formal power series representation $\sum_{k=0}^\infty b_k x^k,$ and a rational function defined by $R_{N,M}=P_N/Q_M,$ where $P_N$ and $Q_M$ are the polynomials of
\begin{align}\label{eqn:FP_1}
P_N(x)=\sum_{n=0}^N p_{n} x^{n}\hbox{ and } Q_M(x)=\sum_{m=0}^M q_{m} x^{m},
\end{align}
respectively, then we say that $R_{N,M}=P_N/Q_M$ is the \textsl{(linear) Pad\'e approximant} of order (N, M) of the formal series that satisfies the condition
\begin{align}\label{eqn:FP_2}
\left(\sum_{n=0}^N p_{n} x^{n}\right)-\left(\sum_{m=0}^M q_{m} x^{m}\right)\left( \sum^{M+N}_{k=0} b_k x^k\right)&=\mathcal{O}(x^{N+M+1}).
\end{align}
Here, $f$ is approximated by $\sum^{M+N}_{k=0} b_k x^k,$. To obtain the approximant $R(N, M),$ we simply calculate the coefficients of polynomials $P_N$ and $Q_M$ by solving the following system of linear equations:
\begin{align}
\sum_{j=0}^M b_{N-j+k}q_k=0,\quad k=1,\ldots,M.\\
\sum_{j=0}^k b_{k-j}q_j=p_k,\quad k=1,\ldots,N.
\end{align}
For this system to be well determined, we usually employ a normalisation by setting, for example, $q_0 = 1.$

If we now consider any piecewise analytic real function $f$ in a finite interval $[a, b]$ with a set of jump locations $\{\zeta_s\}_{s=1}^S\in[a,b]$ that appear in $f$, the complex Fourier series (CFS) representation of the function is defined as
\begin{eqnarray}
f(x)=\mathfrak{Re}\left[\sum_{k=-\infty}^{\infty} b_k e^{i\frac{2\pi}{b-a}k x}\right],\,\hbox{with}\,\, b_k=\frac{1}{b-a}\int_{a}^{b} f(x)e^{-i\frac{ 2\pi }{b-a}kx} dx. \label{FourierS}
\end{eqnarray}
Here, $\mathfrak{Re}$ represents the real part of the function.  As we focus on approximating a real function, we can further obtain
\begin{eqnarray}
f(x)=\mathfrak{Re}\left[2\sum_{k=1}^{\infty} b_k e^{i\frac{2\pi}{b-a}kx}+b_0\right]\label{FourierS_Final}.
\end{eqnarray}
Based on this representation, we denote $z$ as $\exp\left({i\frac{2\pi}{b-a}x}\right)$, and then, we approximate $f$ with a truncated power series of $f_1$ such that
\begin{eqnarray}\label{eqn:CFS_ztransform}
f(x)\approx f_1(z)=\mathfrak{Re}\left[2\sum_{k=1}^{N+M} b_k z^k+b_0\right].
\end{eqnarray}
The transformation $z=\exp\left({i\frac{2\pi}{b-a}x}\right)$ also suggests that the jump location $\varepsilon$  translates into $\exp\left(i\frac{2\pi}{b-a}\zeta\right).$
Based on (\ref{eqn:FP_2}), the Fourier-Pad\'e approximation of $f_1$ comprises the polynomials
\begin{align}\label{eqn:FP}
P_N(z)=Q_M(z)f_1(z)+\mathcal{O}(z^{N+M+1}),	\quad z\rightarrow 0.
\end{align}
However, \cite{Dris_Ben:2001,Dris_Ben:2011} note that this approximant (\ref{eqn:FP}) does not reproduce very well at/around the jump locations of the function, which makes the approximation inaccurate. Therefore, they suggest that every jump $\varepsilon$ can be attributed to a logarithm of the form
\begin{align}
\log\left(1-{z\over \varepsilon}\right)
\end{align}
This logarithmic jump in $f_1$, which is difficult for the Pad\'e approximant to simulate, can be exploited to enhance the approximation process. This is the rationale behind the SFP method introduced
in \cite{Dris_Ben:2001,Dris_Ben:2011}. We modify the Fourier-Pad\'e approximant (\ref{eqn:FP}) to obtain the following condition:
\begin{align}\label{eqn:SFP_1}
P_N(z)+\sum_{s=1}^S L_{N_s}(z)\log\left(1-z/\varepsilon_s \right)=f_1(z)Q_M(z)+\mathcal{O}(z^{U+1}),
\end{align}
where
%
\begin{eqnarray}
\begin{array}{rclrcl}
P_N(z)&=&\sum_{n=0}^N p_{n} z^{n},&Q_M(z)&=&\sum_{m=0}^M q_{m} z^{m}\neq 0,\\
L_{N_s}(z)&=&\sum_{n_s=0}^{N_s}l_{n_s} z^{n_s},& s&=&1,\ldots, S,\\
U&=&N+M+S+\sum_{s=1}^S N_s.&&&
\end{array}
\end{eqnarray}

\section{Financial modelling with L\'evy processes}\label{sec:Fin_Model}
We briefly review option pricing theory in L\'evy-models partly to establish notations. Standard references for this material are \cite{App:2004}, \cite{Con_Tan:2004}, and \cite{Sato:1999}. Throughout this section, we consider that markets are frictionless and have no arbitrage, and we assume that an equivalent martingale measure (EMM) $\mathds{Q}$ is chosen by the market. Moreover, there is a complete filtered probability space $\left(\Omega, \mathcal{F}, \{\mathcal{F}\}_{t\geq 0}, \mathds{Q}\right)$ on which all processes are assumed to live.

We first introduce a stock price process $S=(S_t)_{t\leq 0}$ and assume that it follows an exponential L\'evy process:
\begin{equation}\label{eqn:stckPro}
S_t = S_0 e^{L_t } , \ \ t \geq 0 ,
\end{equation}
where, $S_0\in\mathds{R}^+=(0,\infty)$ is the initial stock price taken as a random variable (rv) independent of $(L_t)_{t\leq 0}$. We limit ourselves to derivatives written on a single risky asset whose log-return we assume to be modelled by a one-dimensional L\'evy process. As usual, we also assume the existence of a risk-free bond earning interest at a constant rate of $r$ and a continuous compounding stock dividend $q$ for all maturities $T>0.$ For a general L\'evy process, the market that consists of the risky asset plus the risk-free bond will be an incomplete market\footnote{Markets are complete when the L\'evy process is a Brownian motion - the classical Black
and Scholes model - or if it is a Poisson process}.

The L\'evy-process $(L_t )_{t \geq 0 } $ is fully determined by its characteristic function that according to the L\'evy--Khinchine theorem, is of the form $\varphi(u):=\mathbb{E }(e^{i u L_t }) = e^{t \phi (u ) } $, with characteristic exponent $\phi (z) $ given by
\begin{align} \label{eqn:LK1}
\phi (u) = i \gamma u - \frac{1 }{2 } \sigma ^2 u^2 + \int_{\mathbb{R } } \, \left( e^{ix u} - 1 - i \chi u \mathbf{1 }_{ \{ |\chi| \leq 1 \} } \right) \nu (\mathrm{d} \chi) .
\end{align}
Here, $\gamma $ and $\sigma $ are real constants with $\sigma \geq 0$, and $\nu $  being a positive measure of $\mathbb{R } $, which is called the L\'evy measure that satisfies the L\'evy-condition $\int_{\mathbb{R } } \, \min (\chi^2 ,1 ) \, \nu (\mathrm{d}\chi) < \infty $. The probabilistic interpretation of $\nu $ is that $\nu (\mathrm{d}\chi) $ gives the expected number of jumps with a size between $\chi$ and $\chi+ \mathrm{d}\chi$, which the process makes between time 0 and 1. The triplet $(\gamma, \sigma, \nu ) $ is called the characteristic triplet or the L\'evy-Khintchine triplet of $(L_t )_{t \geq 0 } .$

We also assume that $\mathds{E}[S_0]\leq 0$ and a recall of (\ref{eqn:stckPro}). Then, we can write
\begin{align}
\mathds{E}[S_t]=\mathds{E}[S_0]\mathds{E}[e^{L_t}]=\mathds{E}[S_0]e^{t\phi(1)},
\end{align}
where $\phi(1)$ is assumed to be finite. For any EMM, $\mathds{Q}$ is a risk-neutral (no-arbitrage) pricing, and the discounted stock price process, $(e^{-(r-q)t}S_t)_{t\geq 0},$ in an equilibrium, with either a complete or an incomplete market, must constitute a martingale. In addition, under the EMM $\mathds{Q}$ measure, the growth rate $\phi(1)$ of the stock price equals the risk-free rate $r > 0$ and $q>0.$

\section{Pricing formulae for European type options}\label{sec:OptionFormu}
In this section, we derive an SFP European option pricing formula. The technique demonstrated is slightly different to the approach in \citet{Chan:2018} as we provide an option pricing curve rather than an individual value.

A European option can be exercised at maturity $T$ of the option. By providing the current log price $x := \log S,$ the strike price of $K$ and the probability density function (PDF) $f$ of a stochastic process, we can express the option price $V(x,K,t)$ starting at time $t$ with its contingent claim that pays out $U(S_T)$ as follows:
\begin{align}\label{eqn:GEquation_1}
V(x,K, t)&=e^{-r (T - t ) } \mathbb{E } (U(S_T,K,T) \vert S_t = e^x )\nonumber\\
&=e^{-r (T - t ) } \mathbb{E } (U(S_te^{X_T-X_t},K,T))\nonumber)\\
&=e^{-r(T-t)}\int_{-\infty}^{+\infty} G(e^{x + \chi-\log K}) f(\chi) d\chi,\quad \chi \in X_T-X_t,
\end{align}
where, $U(S_te^{X_T-X_t},K,T)=G(e^{x + \chi-\log K}).$ By replacing $x+\chi-\log K$ with $y$, we have
\begin{align}
V(x,K,t)&=e^{-r(T-t)}\int_{-\infty}^{+\infty} G(e^{y}) f\left(y-x+\log K \right) \mathrm{d}y\label{eqn:GEquation_Cross}\\
&=e^{-r(T-t)} \int_{-\infty}^{+\infty} G(e^{y}) f^R \left(\tilde{x} - y \right) \mathrm{d}y,\label{eqn:GEquation_Conv}\
\end{align}
where, $\tilde{x}=x-\log K,$ $G(e^{y})$ is the pay-off in the log-price coordinates, and $f^R (\tilde{x}) := f(-\tilde{x}) $ is the reflected function. The expression of (\ref{eqn:GEquation_Cross}) is indeed a cross-correlation integral; however, since we introduce the idea of the reflected function $f^R (\tilde{x}) := f(-\tilde{x})$, we can instead turn (\ref{eqn:GEquation_Cross}) into a convolution integral (\ref{eqn:GEquation_Conv}).

If we consider to approximate $V(x,K,t)$ in a finite interval $[c,d]$ rather than in $[-\infty, \infty],$ such that the choice of $[c,d]$ satisfies the condition of
\begin{align}\label{eqn:charAppx}
\int_{c}^{d} f(\chi) e^{iu\chi} \rm{d}\chi\approx\int_{-\infty}^{+\infty} f(\chi) e^{iu\chi} \rm{d}\chi=\mathds{E}[e^{iu(X_T-X_t)}]:=\varphi(u),
\end{align}
where $\varphi(u)$ is a characteristic function of $X_T-X_t.$, then (\ref{eqn:GEquation_Conv}) becomes
\begin{align}\label{eqn:GEquation_3}
V(x,K, t)&\approx e^{-r(T-t)}\int_{c}^{d} G(e^y)f^R(\tilde{x}-y) \mathrm{d}y.
\end{align}
By using the Fourier transform shift theorem and the CFS expansion shown in (\ref{FourierS_Final}),  we express $f^R\left(\tilde{x}-y\right)$ as
 \begin{align}\label{eqn:CFS_PDF}
\mathfrak{Re}\left[\sum_{k=-\infty}^{+\infty} b_k e^{-i\frac{2\pi}{b-a}ky}\right],
\end{align}
where
\begin{align}
b_k=\frac{1}{d-c}\int_{c}^{d} f(y)e^{-i\frac{ 2\pi }{d-c}ky} \mathrm{d}y\left(e^{i\frac{2\pi}{d-c}k\tilde{x}}\right)\quad\hbox{and}\quad b_0=\frac{1}{d-c}\int_{c}^{d} f(y)\mathrm{d}y.
\end{align}
%
%
Through substitution, we have
\begin{align}\label{eqn:Euro_CFS}
V(x,K,t)=e^{-r(T-t)}\mathfrak{Re}\left[\sum_{k=-\infty}^{+\infty} b_k g_k e^{i\frac{2\pi}{d-c}k\tilde{x}}\right],
\end{align}
where,
\begin{align}
b_k&=\frac{1}{d-c}\int_{c}^{d} f^R(y)e^{-i\frac{ 2\pi }{d-c}ky} \mathrm{d}y\quad\hbox{and}\quad b_0=\frac{1}{d-c}\int_{c}^{d} f^R(y)\mathrm{d}y.\\
g_k&=\int_{c}^{d} G(e^y)e^{-i\frac{ 2\pi }{d-c}ky} \mathrm{d}y\quad\hbox{and}\quad g_0=\int_{c}^{d} G(e^y)\mathrm{d}y.
\end{align}
Because of condition (\ref{eqn:charAppx}), we can approximate $b_k$ and $b_0$ as
\begin{align}\label{eqn:B_k}
\widehat{B}_k:=\frac{1}{d-c}\varphi\left(\frac{2\pi}{d-c}k\right)\quad\hbox{and}\quad \frac{1}{d-c}\widehat{B}_0:=\varphi(0)=1,
\end{align}
respectively.
Furthermore, since we only consider a vanilla call/put in this paper, their payoffs are formulated as
 \begin{align}
U(S_t,K,T)=\begin{cases}
\max\left(e^{x+\chi}-K,0\right)=K\max\left(e^{x+\chi-\log K}-1,0\right):\quad\hbox{(call)} \\
\max\left(K-e^{x+\chi},0\right)=K\max\left(1-e^{x+\chi-\log K},0\right):\quad\hbox{(put)}
\end{cases}.
\end{align}
By considering $y:=x+\chi-\log K$ and applying basis
calculus, we have
\begin{align}\label{eqn:G_k_call}
\widehat{G}_k&=\int_{c}^{d} \max\left(e^{y}-1,0\right) e^{-i\frac{2\pi}{d-c}ky}\mathrm{d}y\nonumber\\
&=\left(\frac{d-c}{d-c-i 2\pi k}\left(e^{(1-i\frac{2\pi}{d-c}k)d}-1\right)+\frac{d-c}{i 2\pi k}\left(e^{-i\frac{2\pi}{d-c}kd}-1\right)\right)
\end{align}
for a call, and similarly, we have
\begin{align}\label{eqn:G_k_put}
\widehat{G}_k&=\int_{c}^{d} \max\left(1-e^{y},0\right)  e^{-i\frac{2\pi}{d-c}ky}\mathrm{d}y\nonumber\\
&=\left(\frac{d-c}{d-c-i 2\pi k}\left(e^{(1-i\frac{2\pi}{d-c}k)c}-1\right)+\frac{d-c}{i 2\pi k}\left(e^{-i\frac{2\pi}{d-c}kc}-1\right)\right)
\end{align}
for a put. Accordingly, we replace $b_k$ with $KG_k$, and the new CFS representation of (\ref{eqn:Euro_CFS}) becomes
\begin{align}\label{eqn:EuroCall_CFStemp}
V(x,K,t):=e^{-r(T-t)}K\mathfrak{Re}\left[\sum_{k=-\infty}^{+\infty} \widehat{B}_k\widehat{G}_k  e^{i\frac{2\pi}{d-c}k\tilde{x}}\right].
\end{align}
To express our final pricing formula with the SFP representation, as we know the pricing formula is a real function, we can transform (\ref{eqn:EuroCall_CFStemp}) into
\begin{align}\label{eqn:temp1}
V(x,K,t):=e^{-r(T-t)}K\mathfrak{Re}\left[2\sum_{k=1}^{\infty} \widehat{B}_k\widehat{G}_k  e^{i\frac{2\pi}{d-c}k\tilde{x}} + \widehat{B}_0\widehat{G}_0\right].
\end{align}
We set $\exp\left({i\frac{2\pi}{d-c} \tilde{x}}\right)$ equal to $z.$ The transformation $z=\exp\left({i\frac{2\pi}{d-c}\tilde{x}}\right)$ maps the interval $[c,d]$ onto the unit circle in $z.$ This change also transforms the jumps $\zeta$ along $f$ into $z$ with the form of $\varepsilon=\exp\left(i\frac{2\pi}{d-c}\zeta\right).$ Finally, by expressing (\ref{eqn:temp1}) with a new variable of $z,$ we have
\begin{align}\label{eqn:temp2}
2\sum_{k=1}^{\infty} \widehat{B}_k\widehat{G}_k z^k + \widehat{B}_0\widehat{G}_0.
\end{align}
By substituting the equation above with $f_1(z)$ in (\ref{eqn:SFP_1}), we obtain the approximant  given by
\begin{align}\label{eqn:EuroCall_SPF1}
P_N(z)+\sum_{s=1}^S L_{N_s}(z)\log\left(1-z/\varepsilon_s \right)=\left(2\sum_{k=1}^{U} \widehat{B}_k\widehat{G}_k z^k + \widehat{B}_0\widehat{G}_0\right)Q_M(z)+\mathcal{O}(z^{U+1})
\end{align}
\begin{align}
\begin{array}{rclrcl}
P_N(z)&=&\sum_{n=0}^N p_{n} z^{n},&Q_M(z)&=&\sum_{m=0}^M q_{m} z^{m}\neq 0,\\
L_{N_s}(z)&=&\sum_{n_s=0}^{N_s}l_{n_s} z^{n_s},& s&=&1,\ldots, S,\\
\varepsilon_s&=&e^{i\frac{2\pi}{d-c}\zeta_s},&U&=&N+M+\sum_{s=1}^S N_s.
\end{array}
\end{align}
Once we can determine the unknown coefficients of $\{p_n\}_{n=0}^N,$ $\{q_m\}_{m=0}^M$ and $\{l_{n_s}\}_{n_s=0}^{N_s}$ in ({\ref{eqn:EuroCall_SPF1}) via the algorithm shown in Appendix \ref{sec:SFP_algorithm} and replace
$$2\sum_{k=1}^{\infty} \widehat{B}_k\widehat{G}_k  e^{i\frac{2\pi}{d-c}k\tilde{x}} + \widehat{B}_0\widehat{G}_0$$
with
$${P_N(z)+\sum_{s=1}^S L_{N_s}(z)\log\left(1-z/\varepsilon_s \right) \over Q_M(z)},\quad z=\exp\left({i\frac{2\pi}{d-c}\tilde{x}}\right),\quad \tilde{x}=x-\log K$$
in ({\ref{eqn:Euro_CFS}), we reach our first SFP representation of a European vanilla option such that
\begin{align}\label{eqn:EuroCall_SPF2}
V(x,K,t):=e^{-r(T-t)}K\mathfrak{Re}\left({P_N(z)+\sum_{s=1}^S L_{N_s}(z)\log\left(1-z/\varepsilon_s \right)\over Q_M(z)}\right).
\end{align}
The pricing formula above can only be applied to compute the option prices with a value of $K$ and a range of $S_t.$ However, in the financial markets, option price quotes always appear with a value of $S_t$ and a range of $K.$ To fit in this financial phenomenon, we modify (\ref{eqn:EuroCall_SPF2}) by using $K=S e^{-\tilde{x}}=e^{x-\tilde{x}}$ so that we obtain the new pricing formula of
\begin{align}
V(x,K,t):=e^{-r(T-t)+x-\tilde{x}}\mathfrak{Re}\left({P_N(z)+\sum_{s=1}^S L_{N_s}(z)\log\left(1-z/\varepsilon_s \right)\over Q_M(z)}\right).
\end{align}

\section{Pricing early-exercise options with the SFP--FCC method} \label{sec:EEOptions}
In this section, we derive option pricing/hedging formulas for early-exercise options by using the SFP--FCC method. We formulate a Bermudan option pricing curve as the first illustration. Then, in the same fashion, we derive the SFP--FCC pricing formulas for the American and discrete-monitored barrier options and their hedging formulas.

The general idea of the SFP--FCC method is first to discretise the lifespan of the options in an equal time step. Then, starting backwards from the maturity to the initial time of the option, we present the option pricing/hedging curve that applies the CFS method at each time step. The accuracy of the CFS method can only be guaranteed by implementing the FCC rules. Finally, once we reach the initial time of the option, the pricing/hedging formula of the option can be constructed by applying the SFP method.

\subsection{Pricing formulae for Bermudan and American options}\label{sec:Bermudan}
We consider $\log S_t:=x_t$ driven by a L\'evy process and a Bermudan option with strike $K$ and maturity $T$ that can be exercised only on a given number of exercise dates $t=t_0 <t_1\leq t_2\leq\ldots t_l\leq t_{l+1}\leq\ldots\leq t_L = T.$ By assuming that the difference between $t_l$ and its successive $t_{l+1}$ is the same, we can write the Bermudan pricing formula for such an option as
\begin{align}\label{eqn:Bermuda_formulae1}
V(x_{t_l},K,t_l)&=\begin{cases} U(e^{x_{t_l}},K,t_l) & l=L,\, t_L=T\\
\max\left(C(x_{t_l},K,t_l), U(e^{x_{t_l}},K,t_l) \right) & l=1,2,3,\dots,L-1\\
C(x_{t_l},K,t_l)&l=0
\end{cases},
\end{align}
where $U(e^{x_{t_l}},K,t_l)$ is the payoff function at $t_l.$ For example, if the payoff function is a call, then $U(e^{x_{t_l}},K,t_l)$ is transformed into $\max\left(e^{x_{t_l}}-K,0\right).$ In (\ref{eqn:Bermuda_formulae1}), $C(x_{t_l},K,t_l)$ at each $t_l$ can be defined as
\begin{align}
C(x_{t_l},K,t_l)&=e^{-r(t_{l+1}-t_l)}\mathbb{E}\left(V(x_{t_{l+1}},K,t_{l+1}) \vert x_{t_l} \right).\\
&=e^{-r(t_{l+1}-t_l)}\int^{+\infty}_{-\infty} V(x_{t_l}+\chi-\log K,t_{l+1})f(\chi)\mathrm{d} \chi,\quad \chi\in X_{t_{l+1}}-X_{t_l}.
\end{align}
Following the algorithm of pricing European options in Section~\ref{sec:OptionFormu}, we set $\tilde{x}_{t_l}=x_{t_l}-\log K,$ replace $\tilde{x}_{t_l}+\chi$ with $y_{t_l}$ and choose $[c,d]$ to satisfy (\ref{eqn:charAppx}). We can transform the equation above as a convolution integral, i.e.,
\begin{align}\label{eqn:int_temp1}
C(x_{t_l},K,t_l)&=e^{-r(t_{l+1}-t_l)}\int^{d}_{c} V(y_{t_l}, t_{l+1})f^R(\tilde{x}_{t_l}-y_{t_l})\mathrm{d} y_{t_l}.
\end{align}

Due to the early-exercise feature of the option, $V(y_{t_l}, t_{l+1})$ is equal to $\max\left(C(y_{t_l},t_{l+1}), U(e^{y_{t_l}},t_{l+1})\right).$ Then, the integral of $C(x_{t_l},K,t_l)$ in (\ref{eqn:int_temp1}) can be split into two parts when
we know the \textsl{early-exercise point}, $x^*_{t_l}$ at $t_l.$ By supposing that we know $x^*_{t_l}$ (we discuss the techniques of finding $x^*_{t_l}$ in Section~\ref{sec:Newton}), we can split the integral, which defines $C(x_{t_l},K,t_l),$ into two parts: one on the interval $[c, x^*_{t_l}]$ and the second on $[x^*_{t_l}, d],$ i.e.,
\begin{align}\label{eqn:C_temp1}
\small
C(x_{t_l},K,t_l)=\begin{cases}\bigintss_{c}^{x^*_{t_l}} C(y_{t_l},t_{l+1}) f^R(\tilde{x}_{t_l}-y_{t_l})\mathrm{d} y_{t_l}+\bigintss_{x^*_{t_l}}^{d}  U(y_{t_l}, t_{l+1})f^R(\tilde{x}_{t_l}-y_{t_l})\mathrm{d} y_{t_l}:\quad\hbox{(call)}\\
\bigintss_{c}^{x^*_{t_l}} U(y_{t_l},t_{l+1}) f^R(\tilde{x}_{t_l}-y_{t_l})\mathrm{d} y_{t_l}+\bigintss_{x^*_{t_l}}^{d}  C(y_{t_l}, t_{l+1})f^R(\tilde{x}_{t_l}-y_{t_l})\mathrm{d} y_{t_l}: \quad\hbox{(put)}\end{cases}.
\end{align}
In (\ref{eqn:C_temp1}), the integral of $$\int U(y_{t_l},t_{l+1})f^R(\tilde{x}_{t_l}-y_{t_l}) \,\mathrm{d} y_{t_l}$$ is clearly the CFS presentation of a European vanilla call or put on $[x^*_{t_l}, d]$ or $[c, x^*_{t_l}]$, respectively, because $ U(y_{t_l},t_{l+1})$ is a payoff, and the CFS representation of $f^R(\tilde{x}_{t_l}-y_{t_l}),$ which is equivalent to (\ref{eqn:CFS_PDF}), is defined as
\begin{align}
f^R(\tilde{x}_{t_l}-y_{t_l})&=\mathfrak{Re}\left[\sum_{\substack{k=-\infty}}^{+\infty} \widehat{B}_k e^{i\frac{2\pi}{d-c}k(-y_{t_l}+\tilde{x}_{t_l})}\right],
\end{align}
where $\widehat{B}_k$ is the same as (\ref{eqn:B_k}). Accordingly, by using the idea of deriving the CFS European option pricing formula in Section~\ref{sec:OptionFormu} and the result of (\ref{eqn:G_k_call}) and (\ref{eqn:G_k_put}),  we can show that
\begin{align}
&\int_{x^*_{t_l}}^d U(y_{t_l},t_{l+1}) \mathfrak{Re}\left[\sum\limits_{\substack{k=-\infty}}^{+\infty} \widehat{B}_k e^{i\frac{2\pi}{d-c}k(-y_{t_l}+\tilde{x}_{t_l})}\right]\nonumber\\
&\mathrm{d} y_{t_l}=K\mathfrak{Re}\left[\sum_{k=-\infty}^{+\infty} \widehat{B}_k\widehat{G}_k[x^*_{t_l},d]  e^{i\frac{2\pi}{d-c}k\tilde{x}_{t_l}}\right]:\label{eqn:U_closedCall}\,\, \hbox{(call)},\\
&\int_{c}^{x^*_{t_l}} U(y_{t_l},t_{l+1}) \mathfrak{Re}\left[\sum\limits_{\substack{k=-\infty}}^{+\infty} \widehat{B}_k e^{i\frac{2\pi}{d-c}k(-y_{t_l}+\tilde{x}_{t_l})}\right]\nonumber\\
&\mathrm{d} y_{t_l}=K\mathfrak{Re}\left[\sum_{k=-\infty}^{+\infty} \widehat{B}_k\widehat{G}_k[c, x^*_{t_l}]  e^{i\frac{2\pi}{d-c}k\tilde{x}_{t_l}}\right]:\label{eqn:U_closedPut}\,\, \hbox{(put)},
\end{align}
where, $\widehat{G}_k[x^*_{t_l},d]$ and $\widehat{G}_k[c, x^*_{t_l}]$ are the closed-form Fourier integrals on $[x^*_{t_l},d]$ and $[c, x^*_{t_l}]$, respectively.

When we compute
\begin{align}\label{eqn:CExpress}
\int C(y_{t_l},t_{l+1}) f^R(\tilde{x}_{t_l}-y_{t_l})\mathrm{d} y_{t_l},
\end{align}
it is not a straightforward case, as $C(y_{t_l},t_{l+1})$ does not have a closed-form expression at $t_{l+1}$. To solve the integral and also yield a higher accuracy of the SFP-FCC method, we first approximate $C(y_{t_l}, t_{l+1})$ with a Chebyshev series since it has a CFS representation in the previous time step. Therefore,
\begin{align}\label{eqn:ChebOpt}
C(y_{t_l},t_{l+1})=C_{cheb}(y_{t_l},t_{l+1}):=\begin{cases}K \sum\limits_{n=1}^{\infty}\alpha_n T_n\circ\psi_{[c, x^*_{t_l}]}(y_{t_l}): \quad\hbox{(call)}\\
K\sum\limits_{n=1}^{\infty}\alpha_n T_n\circ\psi_{[x^*_{t_l},d]}(y_{t_l}): \quad\hbox{(put)}
\end{cases}.
\end{align}
Here, $\alpha_n$ is the $n^{th}$ coefficient, and we also define the composition of $T_k\circ\psi_{[y_k,y_{k+1}]},$ where $\psi_{[y_k,y_{k+1}]}(y_{t_l})=(2y_{t_l}-(y_{k+1}+y_k))/(y_{k+1}-y_k)$ is the linear mapping from $[y_k,y_{k+1}]$ to $[-1,1].$ By substituting (\ref{eqn:ChebOpt}) into (\ref{eqn:CExpress}) and expanding the integral (\ref{eqn:CExpress}), we have
\begin{align}
&\quad\int_c^{x^*_{t_l}} C_{cheb}(y_{t_l},t_{l+1}) f^R (\tilde{x}_{t_l}-y_{t_l}) \mathrm{d} y_{t_l}\nonumber\\
&=K\sum\limits_{k=-\infty}^{+\infty}\sum\limits_{n=1}^{\infty}\widehat{B}_k \alpha_n\left(\int^{x^*_{t_l}}_c T_n\circ\psi_{[c, x^*_{t_l}]}(y_{t_l})e^{-i\frac{2\pi}{d-c}ky_{t_l}} \mathrm{d} y_{t_l}\right)e^{i\frac{2\pi}{d-c}k\tilde{x}_{t_l}}: \quad\hbox{(call)}\label{eqn:chebCCall},\\
&\quad\int_{x^*_{t_l}}^d C_{cheb}(y_{t_l},t_{l+1}) f^R (\tilde{x}_{t_l}-y_{t_l}) \mathrm{d} y_{t_l}\nonumber\\
&=K\sum\limits_{k=-\infty}^{+\infty}\sum\limits_{n=1}^{\infty} \widehat{B}_k \alpha_n\left(\int_{x^*_{t_l}}^d T_n\circ\psi_{[x^*_{t_l},d]}(y_{t_l})e^{-i\frac{2\pi}{d-c}ky_{t_l}} \mathrm{d} y_{t_l}\right)e^{i\frac{2\pi}{d-c}k\tilde{x}_{t_l}}: \quad\hbox{(put)}\label{eqn:chebCPut}.
\end{align}
In the equations above, both integrals of
\begin{align}
\int^{x^*_{t_l}}_c T_n\circ\psi_{[c, x^*_{t_l}]}(y_{t_l})e^{-i\frac{2\pi}{d-c}ky_{t_l}} \mathrm{d} y_{t_l}, \hbox{ and } \int_{x^*_{t_l}}^d T_n\circ\psi_{[x^*_{t_l},d]}(y_{t_l})e^{-i\frac{2\pi}{d-c}ky_{t_l}} \mathrm{d} y_{t_l}
\end{align}
can be simplified into
\begin{align}
\widehat{T}_{n,k}[c, x^*_{t_l}]&:={x^*_{t_l}-c \over 2} e^{-i{d-c \over x^*_{t_l}-c}k\pi}\int_{-1}^{+1}T_n(s)\exp{\left(i\left(-{k (x^*_{t_l}-c)\pi \over d-c}\right)s\right)}\mathrm{d} s\label{eqn:T_expC}:\quad\hbox{(call)}  \\
\intertext{and}
\widehat{T}_{n,k}[x^*_{t_l},d]&:={d-x^*_{t_l}\over 2} e^{-i{d-c \over d-x^*_{t_l}}k\pi}\int_{-1}^{+1}T_n(s)\exp{\left(i\left(-{k (d-x^*_{t_l}) \pi \over d-c}\right)s\right)}\mathrm{d} s:\label{eqn:T_expP}\quad\hbox{(put)},
\end{align}
respectively. We denote $\tilde{k}$ to be equal to either $-{k (x^*_{t_l}-c)\pi \over d-c}$ or $-{k (d-x^*_{t_l}) \pi \over d-c}$ to simplify the mathematical notation in the equations above. Therefore, we have
\begin{align}\label{eqn:FFC_integral}
\int_{-1}^{+1} T_n(s)\exp(i\tilde{k}s)\mathrm{d}s,\quad n\geq 0.
\end{align}
\noindent This integral is not easy to solve numerically because it is highly oscillatory \citep[e.g.,][]{Dom_Gra:2011}. To yield higher accuracy, we apply the FCC rules stated in Appendix \ref{sec:weights} to compute the integral.  By using the final numerical result of (\ref{eqn:FFC_integral}), we can further transform (\ref{eqn:chebCCall}) and (\ref{eqn:chebCPut}) as
\begin{align}
\int_c^{x^*_{t_l}} C_{cheb}(y_{t_l},t_{l+1}) f^R (\tilde{x}_{t_l}-y_{t_l}) \mathrm{d} y_{t_l}=K\sum\limits_{k=-\infty}^{+\infty}\sum\limits_{n=1}^{\infty} \widehat{B}_k \alpha_n\widehat{T}_{n,k}[c, x^*_{t_l}] e^{i\frac{2\pi}{d-c}k\tilde{x}_{t_l}}:\quad\hbox{(call)}\label{eqn:chebCtemp1_call}\\
\int_{x^*_{t_l}}^d C_{cheb}(y_{t_l},t_{l+1}) f^R (\tilde{x}_{t_l}-y_{t_l}) \mathrm{d} y_{t_l}=K\sum\limits_{k=-\infty}^{+\infty}\sum\limits_{n=1}^{\infty} \widehat{B}_k \alpha_n\widehat{T}_{n,k}[x^*_{t_l},d] e^{i\frac{2\pi}{d-c}k\tilde{x}_{t_l}}:\quad\hbox{(put)}\label{eqn:chebCtemp1_put},
\end{align}
respectively. By substituting (\ref{eqn:U_closedCall}), (\ref{eqn:U_closedPut}), (\ref{eqn:chebCtemp1_call}), and (\ref{eqn:chebCtemp1_put}) back into (\ref{eqn:C_temp1}), we can have a CFS representation of $C(x_{t_l},K,t_l)$ such that
\begin{align}\label{eqn:BerCFS_tl}
C(x_{t_l},K,t_l)=e^{-r(t_{l+1}-t_l)}K\begin{cases}\sum\limits_{k=-\infty}^{+\infty}\widehat{B}_k\left(\widehat{G}_k(x^*_{t_l},d) +\sum\limits_{n=1}^{\infty}\alpha_n\widehat{T}_{n,k}[c, x^*_{t_l}]\right)\,e^{i\frac{2\pi}{d-c}k\tilde{x}_{t_l}}:\quad\hbox{(call)}\\
\sum\limits_{k=-\infty}^{+\infty}\widehat{B}_k\left(\widehat{G}_k(c,x^*_{t_l}) +\sum\limits_{n=1}^{\infty}\alpha_n\widehat{T}_{n,k}[x^*_{t_l},d]\right)\,e^{i\frac{2\pi}{d-c}k\tilde{x}_{t_l}}: \quad\hbox{(put)}\end{cases}.
\end{align}
We should notice that the CFS representation above is working at each time step from $t$ and $t_{L-2}.$ However, at $t_{L-1},$ since $t_L=T$ and $V(y_{T}, T)=U(y_{T},T),$  is a payoff function in (\ref{eqn:int_temp1}), we simply have a CFS European pricing formula on $[c,d],$ i.e.,
\begin{align}\label{eqn:BerCFS_L1}
C(x_{t_{L-1}},K,t_{L-1})&=e^{-r(T-t_{L-1})}K\begin{cases}\sum\limits_{k=-\infty}^{+\infty}\widehat{B}_k\widehat{G}_k[0,d]\,e^{i\frac{2\pi}{d-c}k\tilde{x}_{t_{L-1}}}:\quad\hbox{(call)}\\
\sum\limits_{k=-\infty}^{+\infty}\widehat{B}_k\widehat{G}_k[c,0]\,e^{i\frac{2\pi}{d-c}k\tilde{x}_{t_{L-1}}}: \quad\hbox{(put)}\end{cases}.
\end{align}
Finally, to seek an SFP representation of $C(x_t, K, t)$ at time $t,$ we first denote
\begin{align}\label{eqn:mathcal_G}
\widehat{\mathcal{G}}_k=\begin{cases}\widehat{G}_k(x^*_{t_l},d) +\sum\limits_{n=1}^{\infty}\alpha_n\widehat{T}_{n,k}[c, x^*_{t_l}]:\quad\hbox{(call)}\\
\widehat{G}_k(c,x^*_{t_l}) +\sum\limits_{n=1}^{\infty}\alpha_n\widehat{T}_{n,k}[x^*_{t_l},d]:\quad\hbox{(put)}
\end{cases}.
\end{align}
By starting from $T$ using (\ref{eqn:BerCFS_L1}) and then working backwards and recursively using (\ref{eqn:BerCFS_tl}) until $t,$ we can reach
\begin{align}
V(x_t,K,t)=C(x_t,K,t)=e^{-r(t_1-t)}K\left(2\sum\limits_{k=-\infty}^{+\infty}\widehat{B}_k\widehat{\mathcal{G}}_ke^{i\frac{2\pi}{d-c}k\tilde{x}_t}\right).
\end{align}
Then, by following the step proposed in (\ref{eqn:temp1}), we can further infer that
\begin{align}\label{eqn:Ber_CFS}
V(x_t,K,t)=e^{-r(t_1-t)}K\left(2\sum\limits_{k=1}^{\infty}\widehat{B}_k\widehat{\mathcal{G}}_ke^{i\frac{2\pi}{d-c}k\tilde{x}_t}+\widehat{B}_0\widehat{\mathcal{G}}_0\right).
\end{align}
Based on the equation above, we apply all the steps from (\ref{eqn:temp2}) to (\ref{eqn:EuroCall_SPF2}); then, we can reach
\begin{align}\label{eqn:Ber_SFP}
V(x_t,K,t)=e^{-r(t_1-t)}K\mathfrak{Re}\left({P_N(z)+\sum_{s=1}^S L_{N_s}(z)\log\left(1-z/\varepsilon_s \right)\over Q_M(z)}\right),
\end{align}
where $z=\exp\left({i\frac{2\pi}{d-c}\tilde{x}_t}\right)$ and $\tilde{x}_t=x_t-\log K.$

To evaluate American options, one simple approach is to approximate an American option by a Bermudan option with many exercise opportunities $L$ that go into infinity \citep[cf.][]{Fan_Oos:2009b}. An alternative approach is to use a Richardson extrapolation \citep[e.g.][]{Ges_Joh:1984, Cha_Sta:2007}. In this paper, we adapt these two approaches to demonstrate the efficiency of our method. When we use the Richardson extrapolation, we implement the 4-point Richardson extrapolation scheme proposed by \citet{Fan_Oos:2009b}. Accordingly, we have the American option price given by
\begin{align}\label{eqn:amer_extrapolation}
V_{Amer}(L)={1\over 21}\left(64V(2^{L+3})-56V(2^{L+2})+14V(2^{L+1})-V(2^{L})\right),
\end{align}
where $V_{Amer}(L)$ denotes the approximated value of the American option.

\subsection{Early-exercise point using root-finding techniques and a computational algorithm for the Bermudan option}\label{sec:Newton}
In this short section, we combine the SFP--FCC method with root-finding techniques, mainly Newton's method, to find early-exercise points. Newton's method is first proposed in \citet{Fan_Oos:2009b} to find an early-exercise point. This technique can be used when one solves the following equality:
\begin{align}
C(y_{t_l}, t_{l+1})=U(y_{t_l}, t_{l+1}),
\end{align}
which appears in (\ref{eqn:C_temp1}). Therefore, to find $x^*_{t_l},$ we can implement different root-finding techniques, such as the secant method. In this paper, as suggested in \citet{Fan_Oos:2009b}, we instead implement Newton's method (also known as the Newton-Raphson method). The process of this method is repeated as
\begin{align}\label{eqn:Newton}
x_{j+1}=x_j-{U(y_{t_l}, t_{l+1})-C(y_{t_l}, t_{l+1})\over {\partial \over \partial y_{t_l}}U(y_{t_l}, t_{l+1})-{\partial \over \partial y_{t_l}}C(y_{t_l}, t_{l+1})}
\end{align}
over $x_j$ for $j = 1, 2,\ldots$ until a sufficiently accurate value is reached. As we only determine whether $x^*_{t_l}$ lies on $[c,d],$ if not, we set $x^*_{t_l}$ to be equal to the nearest boundary point. In the equation, we start with $x_0$ equal to $x^*_{t_{l+1}},$ the exercise point in the exercise date at $t_{l+1}$, and we also know that at maturity $T,$ $x^*_{T}$ is equal to 0.
In (\ref{eqn:Newton}),
\begin{align}
C(y_{t_l}, t_{l+1})&=e^{-r(t_{l+2}-t_{l+1})}K\Bigg(\mathfrak{Re}\Bigg[2\sum_{k=1}^{\infty}\widehat{B}_k\widehat{\mathcal{G}}_ke^{i\frac{2\pi}{d-c}ky_{t_l}}\Bigg]\Bigg)\label{eqn:CFS_1},\\
{\partial C(y_{t_l}, t_{l+1}) \over \partial y_{t_l}}&=e^{-r(t_{l+2}-t_{l+1})}K\Bigg(\mathfrak{Re}\Bigg[2\sum_{k=1}^{\infty}\left(i\frac{2\pi}{d-c}k\right)\widehat{B}_k\widehat{\mathcal{G}}_ke^{i\frac{2\pi}{d-c}ky_{t_l}}\Bigg]\Bigg).\label{eqn:CFS_2}
\end{align}
Since $C(y_{t_l}, t_{l+1})$ may suffer from the Gibbs phenomenon due to a piecewise continuous PDF. To avoid the phenomenon and achieve a higher accuracy of finding $x^*_{t_l},$ we apply the SFP method to $(\ref{eqn:CFS_1})$ and $(\ref{eqn:CFS_2}).$ To obtain our SFP representation, we first let $z=\exp\left(i{2\pi\over d-c}y_{t_l}\right)$ and then transform all the jumps $\zeta$ into $\varepsilon=\exp\left(i{2\pi\over d-c} \zeta\right)$ in (\ref{eqn:CFS_1}) and (\ref{eqn:CFS_2}). Accordingly, this transforms the CFS representation into the form
\begin{align}
f_1(z)=\begin{cases}
2\sum_{k=1}^{U}\widehat{B}_k\widehat{\mathcal{G}}_kz^k+\widehat{B}_0\widehat{\mathcal{G}}_0,\\
2\sum_{k=1}^{U}\left(i\frac{2\pi}{d-c}k\right)\widehat{B}_k\widehat{\mathcal{G}}_kz^k.\end{cases}
\end{align}
based on the equation above, by using (\ref{eqn:SFP_1}), we can eventually obtain the SFP approximant given by
\begin{align}
P_N(z)\sum_{s=1}^S L_{N_s}(z)\log\left(1-z/\varepsilon_s \right)=f_1(z)Q_M(z)+\mathcal{O}(z^{U+1}).
\end{align}
By applying the approximation algorithm in Appendix \ref{sec:SFP_algorithm}  to determine the coefficients of $P_N,$ $Q_M,$ and $L_{N_s},$ we can obtain the SPF formula for $C(y_{t_l}, t_{l+1})$ and  ${\partial \over \partial y_{t_l}}C(y_{t_l}, t_{l+1})$ with the form
\begin{align}
e^{-r(t_1-t)-x_t}K\mathfrak{Re}\left({P_N(z)+\sum_{s=1}^S L_{N_s}(z)\log\left(1-z/\varepsilon_s \right)\over Q_M(z)}\right).
\end{align}
By combining the root-finding techniques above and summarising Section~\ref{sec:Bermudan}, we present the pseudo-code of our algorithm that computes Bermudan option prices in Algorithm~\ref{algo:Bermudan}.

Finally, we draw our attention to the performance or complexity of the algorithm, $\mathcal{O},$ of the SFP--FCC method. At each time step $t_l$, since we adopt Chebfun \citep{Tre_Dis:2014} to calculate $\alpha_n$ without applying an adaptive process in (\ref{eqn:ChebOpt}), the complexity is $\mathcal{O}(\tilde{N}\log \tilde{N}),$ where $ \tilde{N}$ is the total number of the Chebyshev terms, because Chebfun employs the fast Fourier transfer (FFT) technique, which originated in \citet{Mas_Han:2002}, to calculate $\alpha_n.$ Furthermore, we apply the FFC rules in (\ref{eqn:chebCtemp1_call}) and (\ref{eqn:chebCtemp1_put}), so according to \citet{Dom_Gra:2011}, the complexity of the rules is also $\mathcal{O}(\tilde{N}\log \tilde{N})$ for each complex Fourier term $k$ up to $N.$  Combining the computational complexities above and considering $L$ exercising dates, the total complexity of the SFP--FCC method is $\mathcal{O}((L-1)(N+1)(\tilde{N} \log \tilde{N})).$

\begin{remark}
In (\ref{eqn:CExpress}), we can directly integrate both $C$  and $f^R$ together because they both have a CFS representation with a complex Fourier basis function $e^{-i\frac{2\pi}{d-c}ky_{t_l}};$ however, unfortunately, if we integrate them, our numerical results suggest that less accuracy can be obtained in the SFP framework.
\end{remark}
%
\begin{algorithm}[h]
\caption{Algorithm for computing Bermudan option price $V(x_t, K, t)$ at $t$ by using the SFP-FCC method.}\label{algo:Bermudan}
\KwResult{Bermudan option price $V(x_{t},K,t)$ at time t}
initialisation\;
discretise $[t,T]$ into timesteps $t=t_0, t_1,\ldots, t_l,\ldots, t_L=T$\;
$ t_l=t_{L-1}$\;
compute $C(x_{t_{L-1}},K, t_{L-1})=e^{-r(T-t_{L-1})}K\mathfrak{Re}\left[\sum\limits_{k=-\infty}^{+\infty} \widehat{B}_k\widehat{G}_k  e^{i\frac{2\pi}{d-c}k\tilde{x}_{t_{L-1}}}\right]$ stated in (\ref{eqn:BerCFS_L1})\;
\While{$t_l\neq t$}{
express $C(x_{t_l},K, t_l)$ in the form of (\ref{eqn:C_temp1})\;
find $\tilde{x}^*_{t_l}$ by using the root-finding technique in Section~\ref{sec:Newton}\;
compute $\int U(y_{t_l},t_{l+1})f^R(\tilde{x}_{t_l}-y_{t_l}) \,\mathrm{d} y_{t_l}$ by using the steps from (\ref{eqn:C_temp1}) to (\ref{eqn:U_closedPut})\;
compute $\int C(y_{t_l},t_{l+1})f^R(\tilde{x}_{t_l}-y_{t_l}) \,\mathrm{d} y_{t_l}$ by using the steps from (\ref{eqn:CExpress}) to (\ref{eqn:chebCtemp1_put})\;
express $C(x_{t_l},K, t_l)=e^{-r(t_{l+1}-t_l)}K\mathfrak{Re}\left[\sum\limits_{k=-\infty}^{+\infty} \widehat{B}_k\widehat{\mathcal{G}}_k  e^{i\frac{2\pi}{d-c}k\tilde{x}_{t_l}}\right]$ stated in (\ref{eqn:BerCFS_tl})\;
next $t_l$\;
}
express $C(x_t,K, t)=V(x_t,K, t)=e^{-r(t_1-t)}K\mathfrak{Re}\left({P_N(z)+\sum_{s=1}^S L_{N_s}(z)\log\left(1-z/\varepsilon_s \right)\over Q_M(z)}\right),$ where $z=\exp\left({i\frac{2\pi}{d-c}\tilde{x}_t}\right)$ and $\tilde{x}_t=x_t-\log K,$ by using the steps from (\ref{eqn:Ber_CFS}) to (\ref{eqn:Ber_SFP})\;
\end{algorithm}

\subsection{Pricing formulae for discretely monitored Barrier options}\label{sec:barrier}
A barrier option is an early-exercise option whose payoff depends on the stock price crossing a pre-set barrier level during the option's lifetime. We call the option an up-and-out, knock-out, or down-and-out option when the option's existence fades out after crossing the barrier level. Like European vanilla options, these options can all be written as either put or call contracts that have a pre-determined strike price on an expiration date. In this paper, we only investigate two basic types of barrier options: down-and-out barrier (DO) options and up-and-out barrier (UO) options for the illustrations of our method.
\begin{enumerate}
\item \textsl{Down-and-out barrier (DO) option}: A down-and-out barrier option is an option that can be exercised at a pre-set strike price on an expiration date as long as the stock price that drives the option does not go below a pre-set barrier level during the option's lifetime. As an illustration, if the stock price falls below the barrier, the option is ``knocked-out'' and immediately carries no value.
\item \textsl{Up-and-out barrier (UO) option}: Similar to a down-and-out barrier option, an up-and-out barrier option will be knocked out when the stock price rises above the barrier level during the option's lifetime. Once it is knocked out, the option cannot be exercised at a predetermined strike price on an expiration date.
\end{enumerate}

The structure of discretely monitored barrier options is the same as the structure of Bermudan options. Instead of having a pre-set exercise date and an early-exercise point like Bermudan options, barrier options have a pre-set monitored date and a barrier level. In the case of Bermudan options, when the stock price goes across the early exercise point, a payoff occurs, and the option expires immediately. In the same manner, a barrier option is immediately knocked out when the barrier level is crossed. The barrier level acts exactly the same as the exercise point in Bermudan options. However, in the case of a barrier option without a rebate, no payoff occurs when the barrier level is reached; otherwise, a rebate occurs when a barrier option is knocked out.

In this paper, we only focus on a barrier option without a rebate and use a DO option to illustrate the SFP--FCC method to approximate discretely monitored barrier option prices. Suppose that we have a DO option driven by $S_t$ with a barrier $B$, and  a strike $K$ and a series of monitoring dates $L$: $t=t_0<\ldots<t_l<\ldots < t_L=T;$ the option formulae can be described as
%
\begin{align}\label{eqn:Barrier_formulae1}
V(x_{t_l},K,t_l)&=\begin{cases} U(e^{x_{t_l}},K,t_l)\mathds{1}_{x_{t_l}>\log B } & l=L,\, t_L=T\\
C(x_{t_l},K,t_l)\mathds{1}_{x_{t_l}>\log B}& l=1,\dots,L-1\\
C(x_{t_l},K,t_l)&l=0
\end{cases},
\end{align}
where, $\mathds{1}$ is an indicator function, $ U(e^{x_{t_l}},K,t_l)$ is again either a call or put payoff and
\begin{align}
C(x_{t_l},K,t_l)&=e^{-r(t_{l+1}-t_l)}\int^{d}_{c} V(y_{t_l}, t_{l+1})f^R(\tilde{x}_{t_l}-y_{t_l})\mathrm{d} y_{t_l}.
\end{align}

We follow the steps from (\ref{eqn:int_temp1}) and (\ref{eqn:C_temp1}) in Section~\ref{sec:Bermudan} and replace the exercise point $\tilde{x}^*_{t_l}$ with a scaled log barrier, $\tilde{B}=\log B-\log K.$ Accordingly, we can expand the equation into
\begin{align}\label{eqn:Bar_Ctemp1}
\tiny
C(x_{t_l},K,t_l)=e^{-r(t_{l+1}-t_l)}\left(\int_{\tilde{B}}^{d}  C(y_{t_l}, t_{l+1})f^R(\tilde{x}_{t_l}-y_{t_l})\mathrm{d} y_{t_l}\right).
\end{align}
To compute $\int C(y_{t_l}, t_{l+1}) f^R(\tilde{x}_{t_l}-y_{t_l})\mathrm{d} y_{t_l},$ we follow the steps from (\ref{eqn:CExpress}) to (\ref{eqn:chebCtemp1_put}) in Section~\ref{sec:Bermudan}. We therefore first approximate $C(y_{t_l}, t_{l+1})$ with a Chebyshev series $C_{cheb}(y_{t_l}, t_{l+1}),$ such that
\begin{align}\label{eqn:Bar_Chetemp1}
\int^d_{\tilde{B}} C(y_{t_l}, t_{l+1}) f^R(\tilde{x}_{t_l}-y_{t_l})\mathrm{d} y_{t_l}=K\sum\limits_{k=-\infty}^{+\infty}\sum\limits_{n=1}^{\infty} \widehat{B}_k \alpha_n\widehat{T}_{n,k}[\tilde{B},d] e^{i\frac{2\pi}{d-c}k\tilde{x}_{t_l}}.
\end{align}
By substituting (\ref{eqn:Bar_Chetemp1}) into (\ref{eqn:Bar_Ctemp1}), the CFS representation of $C(x_{t_l},K,t_l)$ can be formulated as
\begin{align}\label{eqn:Bar_Ctemp2}
C(x_{t_l},K,t_l)&=e^{-r(t_{l+1}-t_l)}K\sum\limits_{k=-\infty}^{+\infty}\widehat{B}_k\widehat{\mathcal{G}}_k\,e^{i\frac{2\pi}{d-c}k\tilde{x}_{t_l}},
\end{align}
where $\widehat{\mathcal{G}}_k=\sum\limits_{n=1}^{\infty}\alpha_n\widehat{T}_{n,k}[\tilde{B},d].$ We have a different expression of $\widehat{\mathcal{G}}_k$ in $C(x_{t_{L-1}},K,t_{L-1})$ at $t_{L-1}$ as we do not apply the FCC rules to approximate a payoff function $ U(e^{x_{t_L}},K,t_L)$; therefore, we have
\begin{align}\label{eqn:Bar_CtempL}
C(x_{t_{L-1}},K,t_{L-1})&=e^{-r(t_{l+1}-t_l)}K\sum\limits_{k=-\infty}^{+\infty}\widehat{B}_k\widehat{\mathcal{G}}_k\,e^{i\frac{2\pi}{d-c}k\tilde{x}_{t_l}},
\end{align}
where $\widehat{\mathcal{G}}_k=\widehat{G}_k$ and $\widehat{G}_k$ is either the Fourier transform of a call payoff on $[\tilde{B},d]$ (cf. [\ref{eqn:G_k_call}]) or a put payoff on $[\tilde{B},0]$ (cf. [\ref{eqn:G_k_put}]).
Finally, to have the SFP--FCC pricing formula of the DO barrier option, we work backwards and recursively from $T$ to $t$ by using (\ref{eqn:Bar_Ctemp2}) and (\ref{eqn:Bar_CtempL}) and then approximate $C(x_t,K,t)$ with the SFP approximant at $t$ by applying the steps of (\ref{eqn:Ber_CFS}) and (\ref{eqn:Ber_SFP}) in Section~\ref{sec:Bermudan}. We present the pseudo-code of our algorithm computing DO option prices in Algorithm~\ref{algo:Barrier}.
\begin{algorithm}[h]
\caption{Algorithm for computing discretely monitored DO barrier option price $V(x_t, K, t)$ at time $t$ by using the SFP--FCC method.}\label{algo:Barrier}
\KwResult{discretely monitored barrier option price $V(x_{t},K,t)$ at time $t$}
initialisation\;
discretise $[t,T]$ into timesteps $t=t_0, t_1,\ldots, t_l,\ldots, t_L=T$\;
compute $C(x_{t_{L-1}},K, t_{L-1})=e^{-r(T-t_{L-1})}K\mathfrak{Re}\left[\sum\limits_{k=-\infty}^{+\infty} \widehat{B}_k\widehat{\mathcal{G}}_k  e^{i\frac{2\pi}{d-c}k\tilde{x}_{t_{L-1}}}\right]$ stated in (\ref{eqn:Bar_CtempL})\;
\While{$t_l\neq t$}{
express $C(x_{t_l},K, t_l)$ in the form of (\ref{eqn:Bar_Ctemp2})\;
compute $\int C(y_{t_l},t_{l+1})f^R(\tilde{x}_{t_l}-y_{t_l}) \,\mathrm{d} y_{t_l}$ as stated in (\ref{eqn:Bar_Ctemp1})\;
express $C(x_{t_l},K, t_l)=e^{-r(t_{l+1}-t_l)}K\mathfrak{Re}\left[\sum\limits_{k=-\infty}^{+\infty} \widehat{B}_k\widehat{\mathcal{G}}_k  e^{i\frac{2\pi}{d-c}k\tilde{x}_{t_l}}\right]$ as stated in (\ref{eqn:Bar_Ctemp2})\;
next $t_l$\;
}
express $C(x_t,K, t)=V(x_t,K, t)=e^{-r(t_1-t)}K\mathfrak{Re}\left({P_N(z)+\sum_{s=1}^S L_{N_s}(z)\log\left(1-z/\varepsilon_s \right)\over Q_M(z)}\right),$ where $z=\exp\left({i\frac{2\pi}{d-c}\tilde{x}_t}\right)$ and $\tilde{x}_t=x_t-\log K,$ using the steps from (\ref{eqn:Ber_CFS}) to (\ref{eqn:Ber_SFP})\;
\end{algorithm}

For the UO barrier options, we can modify Algorithm~\ref{algo:Barrier} to compute their prices, but we consider the condition of the option knocked out when the stock price rises above $B,$ i.e.,
\begin{align}
V(x_{t_l},K, t_l)=\begin{cases}  U(e^{x_{t_l}},K,t_l)\mathds{1}_{x_{t_l}<\log B}  & l=L,\, t_L=T\\
C(x_{t_l},K,t_l)\mathds{1}_{x_{t_l}<\log B}& l=1,\dots,L-1\\
C(x_{t_l},K,t_l)&l=0
\end{cases}.
\end{align}

\section{Option Greeks hedging and choice of truncated intervals}\label{sec:greek_truc}
This section is divided into two parts: calculating the option Greeks and choosing truncated intervals. As we have mentioned in \citet{Chan:2018} before, we repeat the deviation of only two option Greeks---Delta and Gamma. Other Greeks, such as Theta, can be derived in a similar fashion; however, depending on the characteristic function, the derivation expression might be rather lengthy. We omit them here, as many terms are repeated. We use the Bermudan option defined in (\ref{eqn:Ber_CFS}) as an illustration to derive the Greeks since the derivation for other option Greeks are the same.

Delta is the first derivative of the value of $V$ of the option with respect to the underlying instrument price S. Therefore, differentiating the CFS expansion of $V$ (\ref{eqn:Ber_CFS}) with respect to $S,$ we have
\begin{align}\label{eqn:CFS_delta}
\Delta_t={\partial V(x_t,K,t) \over \partial S}&={\partial V(x_t,K,t) \over \partial x}{\partial x\over \partial S}\nonumber\\
&=e^{-r(t_1-t)-x_t}K\Bigg(\mathfrak{Re}\Bigg[2\sum_{k=1}^{\infty}\left(i\frac{2\pi}{d-c}k\right)\widehat{B}_k\widehat{\mathcal{G}}_ke^{i\frac{2\pi}{d-c}k\tilde{x}_t}\Bigg]\Bigg).
\end{align}
where $\tilde{x}_t=x_t-\log K.$ Similarly, we can obtain $\Gamma_t$ by differentiating $\Delta_t$ with respect to $S$ such that
\begin{align}\label{eqn:CFS_gamma}
\Gamma_t={\partial^2 V(x_t,K,t) \over \partial S^2}={\partial \Delta_t \over \partial S}={\partial \Delta_t \over \partial x_t}{\partial x_t \over \partial S},
\end{align}
and eventually,
\begin{align}
\Gamma_t&=e^{-r(t_1-t)-2x_t}K\mathfrak{Re}\Bigg[2\sum_{k=1}^{\infty}\left(i\frac{2\pi}{d-c}k\right)\left(i\frac{2\pi}{d-c}k-1\right)\widehat{B}_k\widehat{\mathcal{G}}_ke^{i\frac{2\pi}{d-c}k\tilde{x}_t}\Bigg].\nonumber
\end{align}

To obtain our first SFP representation of $\Delta,$ we first let $z=\exp\left(i{2\pi\over d-c}\tilde{x}_t\right)$ and then transform all the jumps $\zeta$ in $\Delta_t$ into $\varepsilon=\exp\left(i{2\pi\over d-c} \zeta\right)$ in (\ref{eqn:CFS_delta}). Accordingly, this transforms the CFS representation in (\ref{eqn:CFS_delta}) into the form
\begin{align}
f_1(z)=2\sum_{k=1}^{U}\left(i\frac{2\pi}{d-c}k\right)\widehat{B}_k\widehat{\mathcal{G}}_kz^k.
\end{align}
and based on the equation above, by using (\ref{eqn:SFP_1}), we can eventually obtain the SFP approximant given by
\begin{align}
P_N(z)\sum_{s=1}^S L_{N_s}(z)\log\left(1-z/\varepsilon_s \right)=f_1(z)Q_M(z)+\mathcal{O}(z^{U+1}).
\end{align}
By applying the approximation algorithm in Appendix \ref{sec:SFP_algorithm}  to determine the coefficients of $P_N,$ $Q_M,$ and $L_{N_s},$ we can obtain the SPF formula for $\Delta_t$ with the form
\begin{align}
e^{-r(t_1-t)-x_t}K\mathfrak{Re}\left({P_N(z)+\sum_{s=1}^S L_{N_s}(z)\log\left(1-z/\varepsilon_s \right)\over Q_M(z)}\right).
\end{align}
To determine the SFP approximant of $\Gamma_t,$ we follow the same idea of approximating $\Delta_t$ but replace $f_1(z)$ with
\begin{align}
2\sum_{k=1}^{U}\left(i\frac{2\pi}{d-c}k\right)\left(i\frac{2\pi}{d-c}k-1\right)\widehat{B}_k\widehat{\mathcal{G}}_kz^k.
\end{align}

Now we draw our attention to wisely choose a good truncated interval. The choice of the interval $[c,d]$ plays a crucial role in the accuracy of the SFP--FCC method.  A minimum and substantial interval $[c,d]$ can capture most of the mass of a PDF such that our algorithm can, in turn, produce a sensible global spectral convergence rate. We adopt the ideas of \citet{Fan_Oos:2009a} and \citet{Chan:2018} to choose the interval $[c,d]$. In this short section, we show how to construct an interval related to the closed-form formulas of stochastic process cumulants. The idea of using the cumulants was first proposed by \citet{Fan_Oos:2009a} to construct the definite interval $[c,d]$ in (\ref{eqn:charAppx}). Based on their ideas, we have the following expression for $[c,d]$:
\begin{align}\label{eqn:truncInt1}
d&= \left\vert c_1+\tilde{L}\sqrt{c_2+\sqrt{c_4}} \right\vert\nonumber\\
c&=-d,
\end{align}
where $c_1,$ $c_2,$ and $c_4$ are the first, second and fourth cumulants, respectively, of the stochastic process and $\tilde{L}\in[8,12].$ For simple and less-complicated financial models, we also obtain closed-form formulas for $c_1,$ $c_2,$ and $c_4$, which are shown in Table~\ref{table:cumulants} of Appendix \ref{sec:cums}.



\section{Numerical results}\label{sec:results}
The main purpose of this section is to test the accuracy and efficiency of the SFP--FCC method through various numerical tests. This involves evaluating the ability of the method to price any early-exercise options and to exhibit good accuracy even when the PDF is smooth/non-smooth. A number of popular numerical methods are implemented to compare the algorithm in terms of the error convergence and computational time. These methods include the COS method (a Fourier COS series method, \citealp{Fan_Oos:2009a}), the filter-COS method (a COS method with an exponential filter to resolve the Gibbs phenomenon; see \citealp{Ruj_Oos:2013}), the CONV method (an FFT method, \citealp{Lor_Fan:2008}), the FFT--QUAD (a combination of the quadrature and CONV methods; see \citealp{Sul:2005}), and the SWIFT methods (wavelet-based methods; see \citealp{Gra_Oos:2013, Mar:2015, Gra_Oos:2016, Mar_Gra:2017}). When we implement the CONV, we use Simpson's rule for the Fourier integrals to achieve fourth-order accuracy. In the filter-COS method, we use an exponential filter and set the accuracy parameter to $10$ as \cite{Ruj_Oos:2013} report that this filter provides better algebraic convergence than other options. We also set the damping factors of the CONV to 0 for pricing European options.

As the SFP method requests approximating jumps in logarithmic series, we only consider and apply the endpoints $c$ and $d$ as our two known jumps for all non-smooth/smooth PDFs. In all numerical experiments, we use the parameter $U$ to denote the number of terms of the SFP--FCC method, $\tilde{N}$ to denote the number terms of the Chebyshev polynomials and $N$ to denote the number of terms/grid points of the other variables. When we measure the approximation errors of the numerical methods, we use absolute errors, the infinity norm errors $R_\infty$ and the $L_2$ norm errors $R_2$ as the measurement units. A MacBook Pro with a 2.8 GHz Intel Core i7 CPU and two 8 GB DDR SDRAM (cache memory) is used for all experiments. Finally, the code is written in MATLAB, and the codes to implement the COS method and the FFT method, such as the CONV method and the like, are retrieved from \citet{BEN:2015}. In terms of computing the Chebyshev polynomials, we use Chebfun \citep{Tre_Dis:2014} to generate non-adaptive Chebyshev polynomials.

We consider four different test cases based on the following PDFs and other parameters:
\begin{align}
\textbf{VG1}:\, S&=80-120,K = 90,\, \sigma=0.12,\, \theta=-0.14,\,\nu=0.2,\nonumber\\
T&=0.1,\, r=0.1,\, q=0.\\
\textbf{CGMY1}:\, S&=0.5-1.5,\, K=1,\, C=1,\, G=5,\, M=5,\, Y=0.5,\nonumber\\
T&=1,\, r=0.1,\, q=0.0.\\
\textbf{CGMY2}:\,S&=80-120,\, K=100,\, C=4,\, G=50,\, M=60,\, Y=0.7, \nonumber\\
 T&=1, r = 0.05, q=0.02.\\
\textbf{NIG1}:\, S&=100,\, K=80-120,\, \alpha=15,\, \beta=-5,\, \delta=0.5,\, T=1,\nonumber\\
 r&=0.05,\, q=0.02.
\end{align}
In each set of parameters, VG denotes the variance gamma model \citep[e.g.][]{Car_Mad_Cha:1998, Mad_Mil:1991}, CGMY stands for the Carr-German-Maddan-Yor model \citep{Car_Ger_Mad_Yor:2002}, and NIG is short for the normal inverse Gaussian process \citep{Bar:1997}.

Throughout all the numerical tests in this paper, we set $\tilde{L}=8$  in (\ref{eqn:truncInt1}) to obtain an accurate truncated interval for the (filter-)COS, SFP--FCC and SWIFT methods. In the first test, we discuss the behaviour of the error and the stability of the SFP--FCC method if $M$, the number of early-exercise dates, goes to infinity. We also check how the Bermudan option prices converge to their American option counterparts. When $M$ approaches infinity, this leads to $\Delta t$ going to zero and to eventually form a highly peaked PDF. The \textbf{VG1} is chosen for the test because relatively slow convergence was reported for the CONV method for very short maturities in \citet{Lor_Fan:2008}. In the test, the Bermudan call options without paying dividends have the same values as their European counterparts, and the European call reference prices are generated by using the SFP method \citep{Chan:2018}. In Fig.~\ref{fig:VGGraph_TimeErr}, the left-hand side of the graph shows highly peaked PDFs with $\Delta t=0.1$ and $\Delta t=1e^{-05}$, and the right-hand side of the graph demonstrates the logarithm absolute error of the SFP--FCC method. As we gradually increase $M$ from $100$ to $10000$ (equivalent to decrease $\Delta t$ from $0.001$ to $1^{-05}$) and keep both $U=32$ and $\tilde{N}=128$ fixed, the logarithm absolute error stays almost equivalent throughout in the right-hand side of the graph. This indicates that the SFP--FCC method works stably to steadily converge Bermudan option prices to their American option counterparts and yields a spectral convergence rate apart from the jump point. In the next test shown in Fig.~\ref{fig:VGGraph_Compare}, we compare the filter--COS,  CONV, FFT--QUAD methods with the SFP--FCC method for pricing a Bermudan call option with the same input parameters,\textbf{VG1}. In the SFP--FCC method, we set $L$ to $1000$ (equivalent to $\Delta t=1^{-04}$) and gradually increase $U$ in a sequence of $8$ (blue), $16$ (red) and $32$ (yellow), and $\tilde{N}$ is set to be $128$ for the SFP--FCC method. For the rest of the three methods, $N$ is ascended in a sequence of $128$ (blue), $256$ (red) and $512$ (yellow). We compute $401$ Bermudan call option prices in the range of $S$ from 80 to 120 and $K=90.$ Compared with the other methods, we observe that the SFP--FCC method can retain spectral convergence apart from the jump point and yield a higher accuracy than the other methods with fewer summation terms required.

\begin{figure}[h]
\center
\includegraphics[height=7cm,width=14cm]{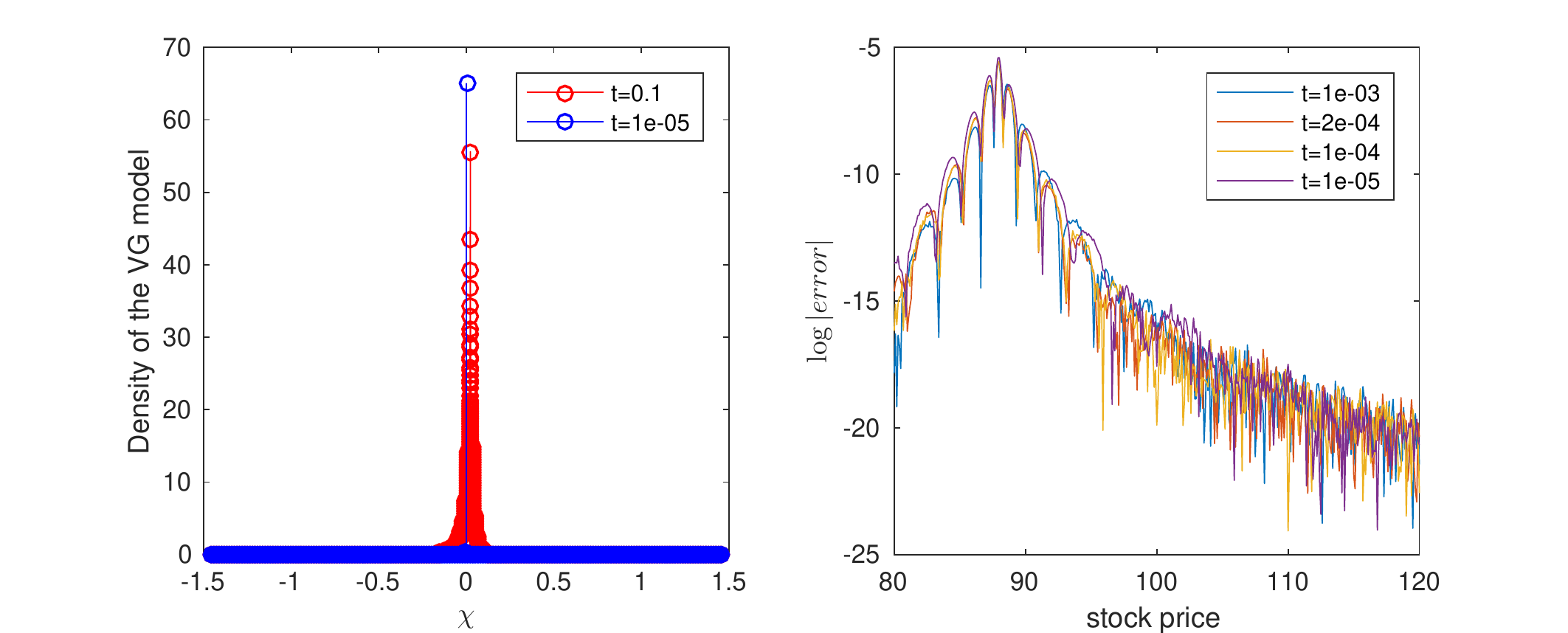}
\setlength{\abovecaptionskip}{1pt}
\caption{Density functions (left) of the VG model and the logarithm absolute errors (right) of the SFP--FCC method with parameters taken from \textbf{VG1}. $L$ is gradually increased in a sequence of $100$ ($\Delta t=1^{-03}$), $500$ ($\Delta t=2^{-04}$), $1000$ ($\Delta t=1^{-04}$) and $10000$ ($\Delta t=1^{-05}$), and both $U$ and $\tilde{N}$ are equal to $32$ and $128$, respectively. $\tilde{L}=8$. $401$ Bermudan call option prices are computed in the range of $S$ from 80 to 120, and $K$ is equal to 90.}\label{fig:VGGraph_TimeErr}
\end{figure}

\begin{figure}[h]
\center
\includegraphics[height=8cm,width=14cm]{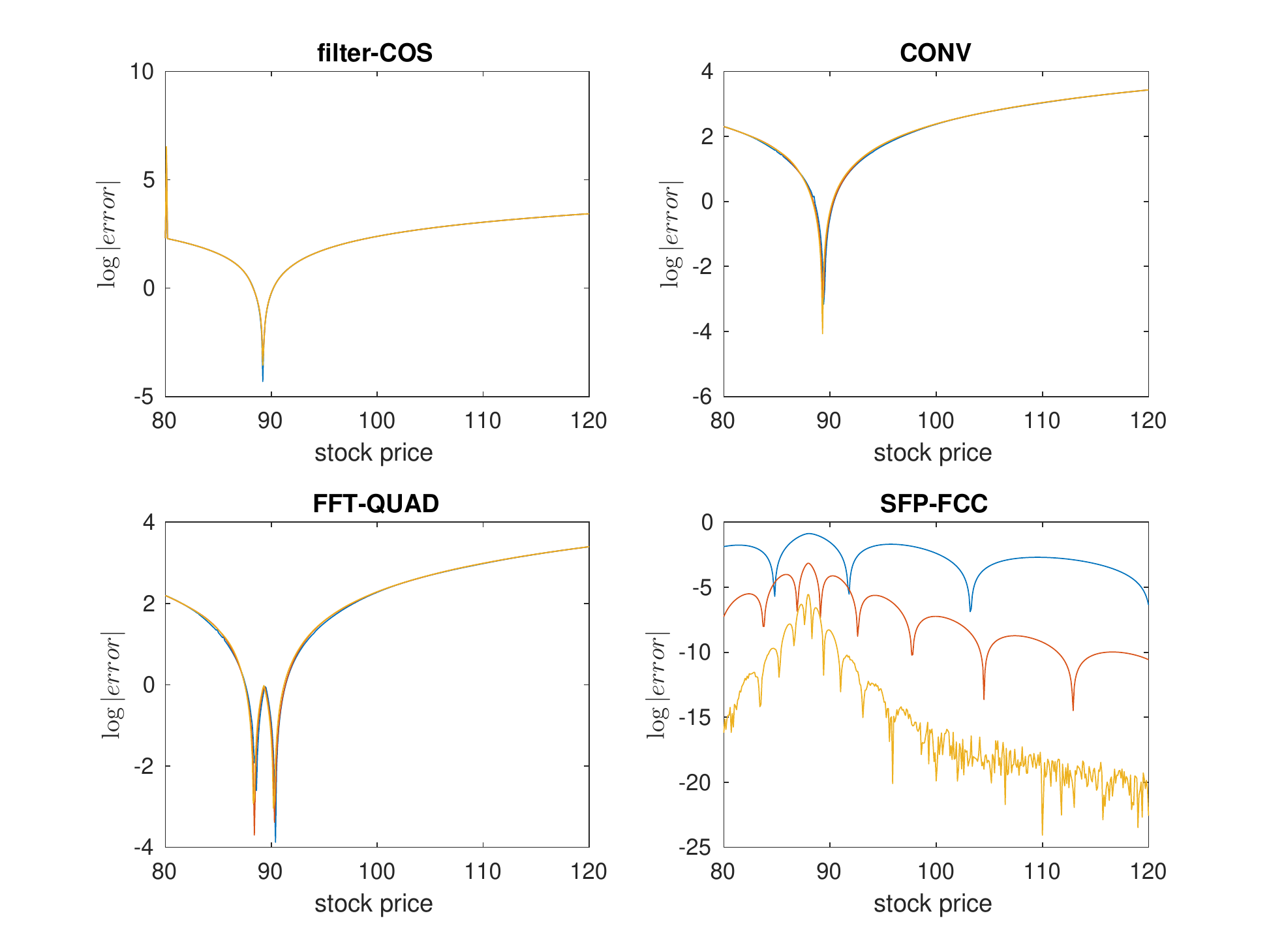}
\setlength{\abovecaptionskip}{1pt}
\caption{Comparison of the filter--COS,  CONV, FFT--QUAD and SFP--FCC methods for pricing a Bermudan call option under the VG model with parameters taken from \textbf{VG1}. $L$ is set to $1000$ (equivalent to $\Delta t=1^{-04}$). $U$ is gradually increased in a sequence of $8$ (blue), $16$ (red) and $32$ (yellow), and $\tilde{N}$ is set to be $128$ for the SFP--FCC method. $N$ is ascended in a sequence of $128$ (blue), $256$ (red) and $512$ (yellow) for the other three methods. $401$ Bermudan call option prices are computed in the range of $S$ from 80 to 120, and $K$ is equal to $90.$ Apart from the jump, spectral convergence is observed in the SFP--FCC method.
}\label{fig:VGGraph_Compare}
\end{figure}

In Table~\ref{table:CGMY_SFPFCC_Amer}, we compare the accuracy of the SFP--FCC method with the COS method in pricing an American put option under the CGMY model after applying the Richardson extrapolation technique (\ref{eqn:amer_extrapolation}) to them. We use \textbf{CGMY1} retrieved from  \citet{Fan_Oos:2009b} for the test. The test itself is a replicate of the same test in \citet[Table 3]{Fan_Oos:2009b}. $14$ reference values are computed by using the CONV method with $N = 4096$ and applying the same extrapolation technique to a range of $S$ from 0.5 to 1.5, and $K$ equals 1. In Table~\ref{table:CGMY_SFPFCC_Amer}, we increase $L$ from $0$ to $3,$ and we can infer that the SFP--FCC method can achieve relatively better accuracy than the COS method with a less total number of $U=256$ and $\tilde{N}=128$ than $N=512$ required. By using the same input parameters of \textbf{CGMY1}, we examine the stability of the SFP--FCC method when $\tilde{N}$ increases in Table~\ref{table:CGMY_SFPFCC_tileN}. We increase $\tilde{N}$ twice from $64$ to $512$ and keep $U=256$ and $L=2$ the same, and both $R_\infty$ and $R_2$ errors first decrease and then level off.

\begin{table}[h]
\caption{Comparison of the COS and SFP--FCC methods for pricing an American put option under the CGMY model with parameters taken from \textbf{CGMY1}; 14 option prices are computed for the CONV method and the COS method in a range of $S$ from 0.5 to 1.5, and $K$ is equal to 1.} \label{table:CGMY_SFPFCC_Amer}
\centering 
\resizebox{\textwidth}{!}{
\begin{tabular}{|c|cccc|ccccc|} 
\hline
\multirow{2}{*}{$L$ in Eq. (\ref{eqn:amer_extrapolation})}&\multirow{2}{*}{$$}&\multicolumn{3}{c|}{\textbf{COS}}&\multicolumn{5}{c|}{\textbf{SFP--FCC}}\\
&$N$&$R_\infty$&$R_2$&Time (sec.)&$U$&$\tilde{N}$&$R_\infty$&$R_2$&Time (sec.)\\
\cline{1-10}
0&512 &  4.182e-02  &  2.717e-01 &0.896&256& 128&    3.180e-02 &   1.797e-01& 0.731\\
1&512  & 1.123e-03   & 9.034e-03 &1.528& 256&128&    1.580e-03 &  9.614e-03 & 1.430\\
2&512   & 2.629e-04  &  2.011e-03 &3.066& 256& 128&    1.659e-05  &  1.011e-04 &3.021\\
3& 512  & 2.667e-05   & 2.021e-04 &6.164& 256& 128 &    1.670e-05  &  1.021e-04& 6.182\\
\hline
\end{tabular}
}
\end{table}

\begin{table}[h]
\caption{Comparison of the $R_\infty$ and $R_2$ errors of the SFP--FCC method for pricing an American put option under the CGMY model with parameters taken from \textbf{CGMY1} when $\tilde{N}$ increases and $L$ and $U$ are kept the same. 14 option prices are computed for the CONV method and the COS method, respectively, in a range of $S$ from 0.5 to 1.5, and $K$ is equal to 1.} \label{table:CGMY_SFPFCC_tileN}
\centering 
\begin{tabular}{|c|ccccc|} 
\hline
\multirow{2}{*}{$L$ in Eq.~(\ref{eqn:amer_extrapolation})}&\multicolumn{5}{c|}{\textbf{SFP--FCC}}\\
&$U$&$\tilde{N}$&$R_\infty$&$R_2$&Time (sec.)\\
\hline
2& 256& 64&    3.180e-03 &  1.114e-02 &  1.530\\
2& 256& 128&    1.659e-05  &  1.011e-04 & 3.021\\
2& 256& 256 &    1.670e-05  &  1.021e-04& 5.282\\
2& 256& 512&    1.670e-05  &  1.021e-04& 10.082\\
\hline
\end{tabular}
\end{table}

In the final two tests, we focus on the comparison of the SFP--FCC method with the SWIFT and COS methods in pricing the UO and DO barrier options, respectively. We set $L$ equal to $12$ and both \textbf{CGMY2} and \textbf{NIG1} are taken from \citet{Fan_Oos:2009b}. All the reference values are generated by using the CONLeg method--the Convolution of Legendre Series \citep{Cha_Hal:2019}. In Tables~\ref{table:CGMY_SFP_UO} and \ref{table:NIG12_SFP_DO}, the difference in the computational time across methods is not large. In Table~\ref{table:CGMY_SFP_UO}, we first compare the accuracy of the SFP--FCC method with the SWIFT method under the CGMY model. In the table, we can see that both methods can reach spectral convergence when we compare 41 UO option prices in the range of $S$ from $80$ to $120,$ $K$ is equal to $100$, and the barrier level, $B$ is set to $120.$ Finally, when pricing the DO barrier options shown in Table~\ref{table:NIG12_SFP_DO} under the NIG model, both methods--COS and SFP--FFC--can obtain spectral convergence when we compare 80 option prices in the range of $K$ from $80$ to $120,$ $S=100$ and $B=80.$ However, the SFP--FCC method can have much lower $R_\infty$ and $R_2$ errors than the COS method when both $N$ and $U$ are doubled. This indicates that the SFP--FCC method is superior to the COS method.

\begin{table}[h]
\caption{Comparison of the SWIFT and SFP--FCC methods for pricing daily-monitored ($L = 12$) UO call and UO put under the CGMY model with parameters taken from \textbf{CGMY2}. 41 option prices are computed in the range of $S$ from 80 to 120, and $K$ is equal to 100. The barrier level $B$ is equal to 120. Spectral convergence is observed in both methods.} \label{table:CGMY_SFP_UO}
\centering 
\resizebox{\textwidth}{!}{
\begin{tabular}{|c|cccc|ccccc|} 
\hline
&\multicolumn{4}{c|}{\textbf{SWIFT}}&\multicolumn{5}{c|}{\textbf{SFP--FCC}}\\
&$scale$&$R_\infty$&$R_2$&Time (sec.)&$U$&$\tilde{N}$&$R_\infty$&$R_2$&Time (sec.) \\
\hline
\multirow{ 5}{*}{UO Call}
&2&    6.419e-01 &   2.522   & 0.208& 8& 128&   3.439e-01 &   8.022e-01 &   0.512\\
&3&    3.344e-02  &  1.391e-01   & 0.256& 16& 128&  6.114e-02   & 2.398e-01   & 0.856\\
&4&    6.710e-04  &  3.231e-03  &  0.324& 32& 128&    1.220e-04  &  4.568e-04  &  0.882\\
&5&    1.287e-07  &  4.560e-06  &  0.451& 64& 128&    3.187e-09  &  1.260e-08   &  0.911\\
&6&    1.561e-12   & 4.850e-12 &   0.761& 128& 128&    1.769e-12 &   5.050e-12  &  1.071\\
\hline
\multirow{ 5}{*}{UO Put}& 2&   1.313&  7.307&    0.206& 8 &128 &    3.353e-01   & 9.707e-01 &    0.123\\
&3&    2.115e-02&    5.742e-02  &  0.264 &16&128   &1.185e-02 &   4.842e-02 &   0.251\\
&4&    5.613e-03 &   2.964e-02   & 0.336 &32&128   &4.663e-05 &   1.964e-04  &  0.321\\
&5&   7.178e-07 &   3.721e-06   & 0.472 & 64 &128   &6.078e-11&    2.724e-10   & 0.425\\
&6&    2.021e-12 &   8.234e-12  &  0.761 & 128 &128 &  1.825e-13&    7.825e-13 &   0.543\\
\hline
\end{tabular}
}
\end{table}

\begin{table}[h]
\caption{Comparison of the COS and SFP--FCC methods for pricing daily-monitored ($L = 12$) DO call and DO put under the NIG model with parameters taken from \textbf{NIG1}. 80 option prices are computed in the range of $K$ from 80 to 120, and $S$ is equal to 100. The barrier level $B$ is equal to 80. Spectral convergence is observed in both methods.} \label{table:NIG12_SFP_DO}
\centering 
\resizebox{\textwidth}{!}{
\begin{tabular}{|c|cccc|ccccc|} 
\hline
&\multicolumn{4}{c|}{\textbf{COS}}&\multicolumn{5}{c|}{\textbf{SFP--FCC}}\\
&$N$&$R_\infty$&$R_2$&Time (sec.)&$U$&$\tilde{N}$&$R_\infty$&$R_2$&Time (sec.) \\
\hline
\multirow{ 4}{*}{DO Call}
&64&1.965e-02&5.741e-02&0.691&64&256&2.837e-03&1.382e-02&0.551\\
&128&1.571e-03&4.244e-03&0.876&128&256&2.905e-05&1.364e-04&0.651\\
&256&1.532e-05&4.138e-05&1.181&256&256&6.871e-08&1.418e-07&0.761\\
&512&3.29e-09&7.867e-09&1.591&512&256&5.351e-10&3.285e-09&1.282\\
\hline
\multirow{ 4}{*}{DO Put} &64 &   4.212e-02   & 1.246e-01   & 0.681 &64 &256   &3.104e-04  &  1.179e-03 &   0.701\\
&128 &  2.632e-03   & 7.166e-03   & 0.712&128& 256  & 1.479e-05 &   8.387e-05&    0.822\\
&256  &  2.811e-05   & 7.358e-05  &  1.060&256 &256   & 2.566e-09  &  1.469e-08 &   0.981\\
&512 &    5.705e-09   & 1.326e-08 &   1.460 &512 &256  &6.377e-13   & 9.154e-13  &  1.350\\
\hline
\end{tabular}
}
\end{table}
\section{Conclusions}\label{sec:conclusion}
We have generalised the SFP option pricing method, based on a singular Fourier--Pad\'e series, to price and hedge early-exercise options--Bermudan, American and discretely-monitored barrier options. We call the new method SFP--FCC, as we incorporate the SFP method with the Filon--Clenshaw--Curtis (FCC) rules. The main advantages of the SFP--FCC method are its ability to return the price and Greeks as a function defined on a prescribed interval rather than just point values and its ability to retain spectral convergence under any process with a (piecewise) continuous PDF. The complexity of the new method is $\mathcal{O}((L-1)(N+1)(\tilde{N} \log \tilde{N}) )$, and the method itself is shown to be favourable to existing popular techniques in all numerical experiments.

Future research on the method will aim to prove theoretically spectral convergence for early-exercise options and extend the method to price options with path-dependant features under the (time-changed) L\'evy process or (rough) stochastic volatility. Research in this direction is already underway and will be presented in a forthcoming manuscript.

\appendix
\setcounter{table}{0}
\setcounter{figure}{0}
\section{Computation of the singular Fourier-Pad\'e coefficients}\label{sec:SFP_algorithm}
The approach to computing the polynomial coefficients needed in the SFP method is fairly straightforward. To demonstrate the algorithm, we focus on a simple case where the option pricing and Greeks formulae are infinitely smooth apart from the jumps located at the endpoints $c$ and $d.$ As we consider $z=\exp\left(i{2\pi\over d-c}\tilde{x}\right)$ in either the option pricing formula or the Greeks formula, the jump of $c$ and $d$ in the z-plane is $-$1. For the sake of simplicity, we denote $f_1(z)$ as the CFS representation of any European-style pricing formula or its option Greeks formula. With some superscripts dropped for clarity and knowing that $s=1,$ in (\ref{eqn:SFP_1}), we have
\begin{align}\label{eqn:SFP_al1}
 P_N(z)+ L_{N_1}(z)\log\left(1-{z\over \varepsilon_1} \right)=f_1(z)Q_M(z)+\mathcal{O}(z^{U+1}),
\end{align}
where $N+M+N_1=U.$ Both  $L_{N_1}$ and $f_1(z)$ have Taylor series and CFS expansions, respectively, to determine U; therefore, their expansions are
\begin{align}
\log\left(1-{z\over \varepsilon_s} \right)&=\sum_{k=1}^{U}-{z^k\over \varepsilon_1^k}+0\\
f_1(z)&=2\sum_{k=1}^{U}\widehat{B}_k \widehat{G}_k z^k+ \widehat{B}_0\widehat{G}_0.
\end{align}
Our goal is to derive a linear system for the unknown polynomial coefficients. Note that $Q_M(z)$ and $L_{N_1}(z)$ are determined only by terms of order greater than $N$. Accordingly, we seek a linear solution to
\begin{align}\label{eqn:SPFMatrix_1}
\begin{bmatrix}
\widehat{B}\widehat{G} &-L \\
\end{bmatrix}
\begin{bmatrix}
\mathbf{q}\\ \mathbf{l}
\end{bmatrix}
=\mathbf{0}.
\end{align}
Here, $\widehat{B}\widehat{G}$ is the $(M+N_1+1)\times(M+1)$ Toeplitz matrix
\begin{align}
\begin{bmatrix}
\widehat{B}_{\frac{U}{2}+1}\widehat{G}_{\frac{U}{2}+1}&\widehat{B}_{\frac{U}{2}}\widehat{G}_{\frac{U}{2}}& \cdots &\widehat{B}_1\widehat{G}_1 \\
\widehat{B}_{\frac{U}{2}+2}\widehat{G}_{\frac{U}{2}+2} &\widehat{B}_{\frac{U}{2}+1}\widehat{G}_{\frac{U}{2}+1}  & \ddots & \widehat{B}_2\widehat{G}_2\\
\vdots&\vdots &\ddots &\vdots\\
\widehat{B}_U\widehat{G}_U & \widehat{B}_{U-1}\widehat{G}_{U-1} &\cdots & \widehat{B}_{\frac{U}{2}}\widehat{G}_{\frac{U}{2}},
\end{bmatrix}
\end{align}
and L is the $(M + N_1 + 1)\times (N_1 + 1)$ matrix defined similarly by using the Taylor coefficients of log(1+z).
The vectors $\mathbf{q}=\{q_m\}_{m=0}^{M}$ and $\mathbf{l}=\{l_{n_1}\}_{{n_1}=0}^{N_1}$ hold the unknown polynomial coefficients in order of increasing degree.
As the column dimension of the matrix in (\ref{eqn:SPFMatrix_1}) is one greater than its row dimension, we can conclude that there is one nonzero solution to (\ref{eqn:SPFMatrix_1}). In many cases, this can be made into a square system by choosing, for example, $q_0 = 1$. However, if one does not want to assume that any particular coefficient is nonzero, one can solve (\ref{eqn:SPFMatrix_1}) by a singular value decomposition. Finally, the unknown coefficients of $\mathbf{p}=\{p_n\}_{n=1}^N$ can be obtained by multiplication through the following matrix system:
\begin{align}
\mathbf{p}
=\begin{bmatrix}
\widehat{B}_0\widehat{G}_0 &  &   &   \\
\widehat{B}_1\widehat{G}_1  & \widehat{B}_0\widehat{G}_0 &  & \\
\vdots&\ddots &\ddots &\\
\widehat{B}_{{U\over 2}}\widehat{G}_{{U\over 2}}  & \cdots &\cdots& \widehat{B}_0\widehat{G}_0
\end{bmatrix}
\mathbf{q}
-
\begin{bmatrix}
l_0 &  &   &   \\
l_1 & l_0 &  & \\
\vdots&\ddots &\ddots &\\
l_{{U\over 2}}  & \cdots &\cdots&\ l_0
\end{bmatrix}
\mathbf{l}.
\end{align}
If there is more than one jump location in the option pricing/Greeks curve (\ref{eqn:SFP_al1}), this suggests the following modification of the equation:

\begin{align}
 P_N(z)+ L_{N_1}(z)\log\left(1-{z\over \varepsilon_1} \right)+\ldots+ L_{N_s}(z)\log\left(1-{z\over \varepsilon_S} \right)=f_1(z)Q_M(z)+\mathcal{O}(z^{U+1}).
\end{align}
Accordingly, we have to modify (\ref{eqn:SPFMatrix_1}) to produce a new $L$ matrix and a vector of coefficients for each location to reflect the changes. According to \cite{Dris_Ben:2001, Dris_Ben:2011}, there is no rigorous optimal formula for choosing the degrees $M,$ $N,$ and $N_1,$\ldots,$N_s.$ Because the denominator polynomial $Q_M$ is shared, we allow $M$ to be the largest, with the others being equal as far as possible. For the case of just one jump location, taking $N$ at roughly $40\%$ of the total available degrees of freedom seems to work well. Experiments suggest that these choices can affect the observed accuracy, occasionally by as much as an order of magnitude, but on average, there is little variation within a broad range of choices.


\section{Locating jumps in probability density functions}\label{sec:sing}
Many PDFs (cf. Fig.~\ref{fig:VGsingularity}) of interest are not smooth but piecewise smooth. If the locations of all jumps are not known in advance in the PDFs, we can also use Fourier-Pad\'e ideas \citep[cf.][]{Dris_Ben:2011,Chan:2018} to estimate the locations of jumps sufficiently well to allow good reconstruction nearly everywhere in the interval $[c,d]$.

\begin{figure}
\center
\includegraphics[height=5cm,width=10cm]{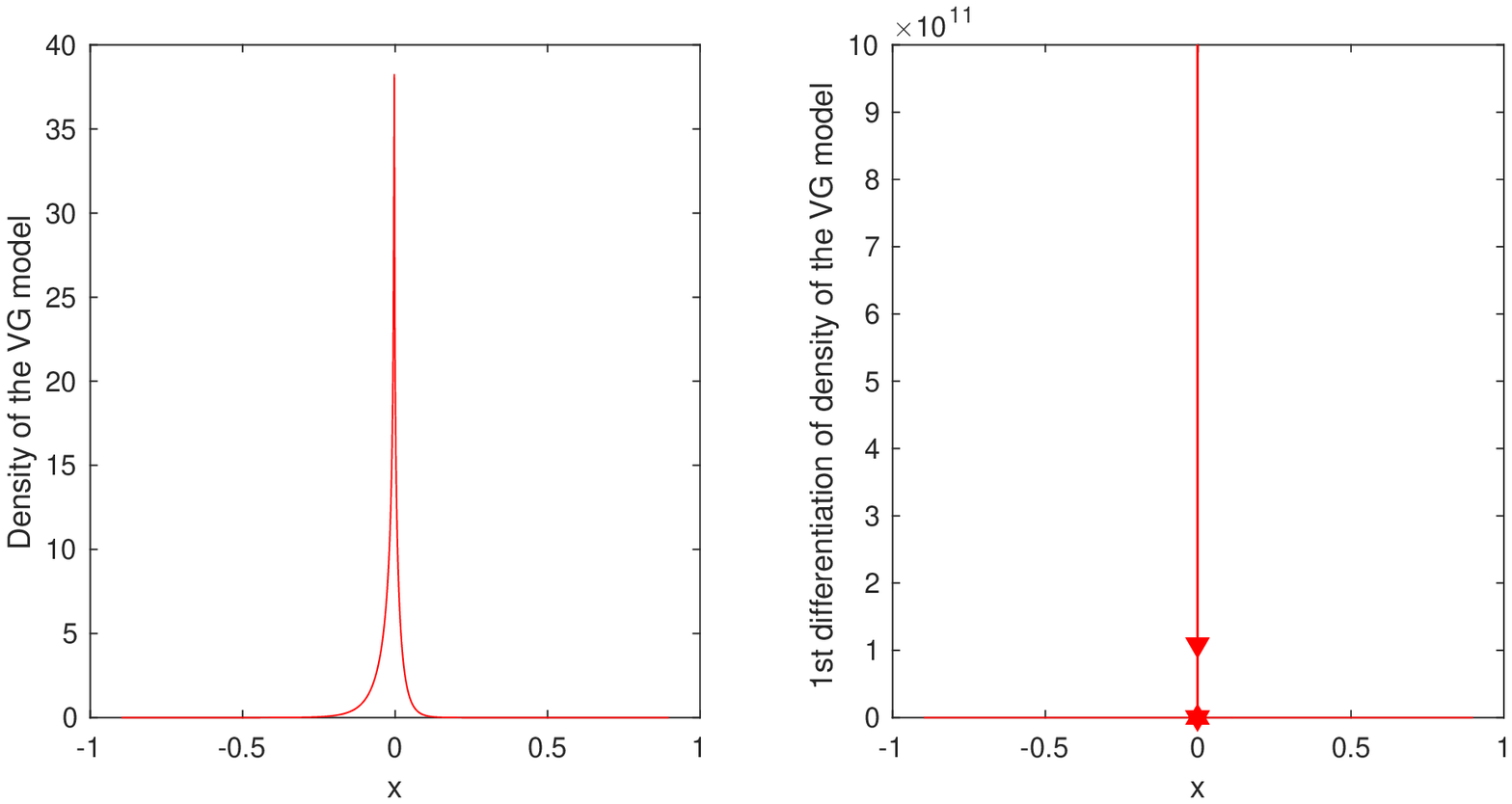}
\setlength{\abovecaptionskip}{1pt}
\caption{Density functions (left) of the VG model and its first derivative (right). The parameters are taken from \textbf{VG1}.}
\label{fig:VGsingularity}
\end{figure}

Here, $g$ is approximated by $\sum^{M+N}_{k=0} b_k x^k,$. To obtain the approximant $R(N, M),$ we simply calculate the coefficients of polynomials $P_N$ and $Q_M$ by solving a system of linear equations. To obtain $\{q_m\}_{m=0}^M,$ we first normalise $q_0 = 1$ to ensure that the system is well determined and has a unique solution in (\ref{eqn:FP_2}). Then, we consider the coefficients for $x^{N+1},\ldots , x^{M+N},$ and we can yield a Toeplitz*\footnote{A Toeplitz matrix or diagonal-constant matrix is an invertible matrix in which each descending diagonal from left to right is constant.} linear system:
\begin{align}
\begin{bmatrix}
b_{N+1} & b_{N} & b_{N-1} & \cdots & b_{N+1-M}\\
b_{N+2} & b_{N+1} & b_{N} & \ddots & b_{N+2-M}\\
\vdots&\ddots &\ddots &\ddots &\vdots\\
b_{N+M} & \cdots &b_{N+2} & b_{N+1} & b_{N}
\end{bmatrix}
\begin{bmatrix}
q_0\\q_1\\ \vdots\\ q_M
\end{bmatrix}
=0.
\end{align}

Once $\{q_m\}_{m=0}^M$ is known, $\{p_n\}_{n=0}^N$ is found through the terms of
order N and less in (\ref{eqn:FP_2}). This yields $\underline{p} = B\underline{q}$, where $b_{ij} = b_{i-j}$. For example, if
$N=M,$ one obtains
\begin{align}
\begin{bmatrix}
p_0\\p_1\\ \vdots\\ p_N
\end{bmatrix}=\begin{bmatrix}
b_{0} & & & \\
b_{1} & b_{0} & & \\
\vdots&\ddots &\ddots &\\
b_{N} & \cdots & b_{1} & b_{0}
\end{bmatrix}
\begin{bmatrix}
q_0\\q_1\\ \vdots\\ q_M
\end{bmatrix}.
\end{align}

Now, assuming $g$ is a PDF, to find the jumps in $g$ and to express $g$ in a Fourier-Pad\'e series, we first express $g$ with the CFS representation:

\begin{align}\label{eqn:CFS_PDF_temp2}
\mathfrak{Re}\left[2\sum_{k=1}^{\infty} \varphi\left(\frac{2\pi}{d-c}k\right) e^{-i\frac{2\pi}{d-c}kx}+\varphi\left(0\right)\right].
\end{align}
Then, we can differentiate (\ref{eqn:CFS_PDF_temp2}) with respect to $x$ to obtain

\begin{align}
\mathfrak{Re}\left[2\sum_{k=1}^{\infty}-\left(i\frac{2\pi}{d-c}k\right) \varphi\left(\frac{2\pi}{d-c}k\right) e^{-i\frac{2\pi}{d-c}kx}\right].
\end{align}
Finally, we let $z=\exp\left({i\frac{2\pi}{d-c}x}\right)$ in the two equations above, and they are ready for the Fourier-Pad\'e approximation. In general, when the PDF has a jump, the sharp-peaked jump point will have an enormously large value after differentiation. In other words, Fig.~\ref{fig:VGsingularity} is a graphical illustration of the outlooks of the PDF (left) and the first derivative (right) of the VG model after the Fourier-Pad\'e approximation. In the figure, we can see that the non-smooth PDF with a jump can produce a value of $10\times 10^{11}$ at the jump point after the first derivative.

\section{Accurate computation of the weights}\label{sec:weights}
We adopt \citet{Dom_Gra:2011}s' algorithm to compute
\begin{align}\label{eqn:w_nFCC}
w_n(\tilde{k}):=\int_{-1}^{+1} T_n(s)\exp(i\tilde{k}s)\mathrm{d}s,\quad n\geq 0.
\end{align}
For the sake of clear mathematical notations, finally, we assume the total number of a Chebyshev series as described in (\ref{eqn:w_nFCC}), which is $N$ in this section.
\subsection{Algorithm: for $n\leq N \leq \tilde{k}$ (first phase)}
First, based on the idea of $U_n=1/(n+1)T_{n+1}'$ \citep[cf.][Eq. (22.5.8)]{Abr_Ste:1965}, where $U_n$ is the $n$th Chebyshev polynomial of the second kind, we can see that
\begin{align}
\rho_n(\tilde{k}):=\int_{-1}^{+1} U_{n-1}(s)\exp(i\tilde{k}s)\mathrm{d}s={1\over n}\int_{-1}^{+1} T_n'(s)\exp(i\tilde{k}s)\mathrm{d}s.
\end{align}
Then, according to \citet[Section 4]{Dom_Gra:2011}, their computation algorithm leads to
\begin{align}
w_n(\tilde{k}):=\gamma_n(\tilde{k})-{n\over ik}\rho_n(\tilde{k}),\quad n\geq 1,\quad w_0(\tilde{k}):=\gamma_0(\tilde{k}).
\end{align}
Here,
\begin{align}
\gamma_n(\tilde{k})=\begin{cases}{2\sin \tilde{k}\over \tilde{k}}&\quad\hbox{for even $n$}\\{2\cos \tilde{k}\over \tilde{k}}&\quad\hbox{for odd $n$}\end{cases},\quad \gamma_0(\tilde{k})={1\over i\tilde{k}}\left(\exp(i\tilde{k})-\exp(-i\tilde{k})\right), \label{eqn:gamma1}
\end{align}
and $\rho_n(\tilde{k})$ can be determined based on the recurrence relationship,
\begin{align}
2\gamma_n(\tilde{k})-{2n\over ik}\rho_n(\tilde{k})=\rho_{n+1}(\tilde{k})-\rho_{n-1}(\tilde{k}),\quad n\geq 2, \label{eqn:rho1}
\intertext{with}
\rho_0(\tilde{k}):=\gamma_0(\tilde{k}) \hbox{ and }\rho_2(\tilde{k}):=2\gamma_1(\tilde{k})-{2\over ik}\gamma_0(\tilde{k}), \label{eqn:rho2}
\end{align}
If $n\leq N\leq \tilde{k},$ by using (\ref{eqn:gamma1}) for computing $\gamma_n(\tilde{k})$ and (\ref{eqn:rho1}) and (\ref{eqn:rho2}) as a forward recurrence for $\rho_n(\tilde{k}),$ we can stably obtain a vector of $\{w_n(\tilde{k})\}_{n=0}^N.$ We summarise the computation in Algorithm \ref{algo_w1}. According to \citet[Theorem 5.1 and Corollary 5.2]{Dom_Gra:2011}, the stability for $n\leq N\leq \tilde{k}$ is proofed. However, the algorithm becomes unstable when $n\geq \tilde{k}$ and $n\leq \tilde{k}\leq N.$
\begin{algorithm}[h]
\caption{Algorithm: for $n\leq N\leq \tilde{k}$ (first phase)}\label{algo_w1}
\begin{algorithmic}[1]
\State Compute
\begin{align}
&\rho_1(\tilde{k}):=\gamma_0(\tilde{k}),\\
&\rho_2(\tilde{k}):=2\gamma_1(\tilde{k})-{2\over i \tilde{k}}\gamma_0(\tilde{k}),\\
&\rho_{n+1}(\tilde{k}):=2\gamma_n(\tilde{k})-{2\over i \tilde{k}}\gamma_n(\tilde{k})+\rho_{n-1}(\tilde{k}),\quad n=2,\ldots,N-1,\quad N\leq \tilde{k}.
\end{align}
\State Set
\begin{align}
w_n(\tilde{k}):=\gamma_n(\tilde{k})-{n\over ik}\rho_n(\tilde{k}),\quad w_0(\tilde{k}):=\gamma_0(\tilde{k}),\quad n=1, 2,\ldots,N,\quad N\leq \tilde{k}
\end{align}
 \end{algorithmic}
\end{algorithm}
\subsection{Algorithm: for $n\leq \tilde{k}<N$ (second phase)}
According to \citet{Dom_Gra:2011}s' algorithm, if $n\leq \tilde{k}<N,$ we must modify Algorithm~\ref{algo_w1}. In this case, we introduce the integers $n_0=\ceil[\big]{\tilde{k}},$ the ceiling function mapping $\tilde{k}$ to the least integer greater than or equal to $\tilde{k},$ and $M\geq n_0,$ the tridiagonal matrix and the right-hand side vector
\begin{align}\label{eqn:AMatrix}
A_{M}(\tilde{k})\boldsymbol{\rho}_M(\tilde{k})= \mathbf{b}_M(\tilde{k}),
\end{align}
where
\begin{align}
A_{M}(\tilde{k}) = \begin{bmatrix}
    {2n_0\over i\tilde{k}} & 1 &    &      &    \\
   -1   & {2(n_0+1)\over i\tilde{k}}        & 1   &     &   \\
      &      -1   & {2(n_0+2)\over i\tilde{k}}  & 1    &   \\
      &         &     \ddots      &     \ddots      & \ddots   \\
      &         &           &         -1  & {2(2M-1)\over i\tilde{k}}
  \end{bmatrix},\,
   \mathbf{b}_M(\tilde{k}):=\begin{bmatrix}
  2\gamma_{n_0}(\tilde{k})+\rho_{n_0-1}(\tilde{k})\\
    2\gamma_{n_0+1}(\tilde{k})\\
  2\gamma_{n_0+2}(\tilde{k})\\
  \vdots\\
 2\gamma_{2M-1}(\tilde{k})+\rho_{2M}(\tilde{k})
\end{bmatrix}
\end{align}
\begin{align}
\boldsymbol{\rho}_M(\tilde{k}):=\left[\rho_{n_0}(\tilde{k})\quad \rho_{n_0+1}(\tilde{k})\quad \rho_{n_0+2}(\tilde{k})\quad \cdots \quad \rho_{2M-1}(\tilde{k})\right]^{\rm T}.
\end{align}
Since $A_M(\tilde{k})$ is a tridiagonal matrix, we can use Oliver's algorithm \citep{Oli:1967}, proposed by \citet{Dom_Gra:2011}, to solve (\ref{eqn:AMatrix}) to obtain $\boldsymbol{\rho}_M(\tilde{k}).$  The coefficients $\gamma_n(\tilde{k})$ and $\rho_{n_0-1}(\tilde{k})$ can be obtained by Algorithm~\ref{algo_w1}. The value of $\rho_{2M}(\tilde{k})$ is a priori unknown, but if we take $2M$ sufficiently large, we can approximate it accurately by using an asymptotic expansion as shown in the next algorithm.
\subsection{Algorithm: for $\tilde{k}<n<N$ (thrid phase)}
According to \citet[Theorem 3.1]{Dom_Gra:2011}, if $M$ is sufficiently large, then we can compute the asymptotic expansion of $\rho_{2M}(\tilde{k})$ with a formula of
\begin{align}\label{eqn:asym_rho}
2i\left(\sum\limits_{r=0}^J(-1)^r p_{2r}(0)\sin \tilde{k}+\sum\limits_{r=0}^J(-1)^r p_{2r+1}(0)\cos \tilde{k}\right)+R_J(M,k),
\end{align}
where the coefficients are defined as
\begin{align}
p_0(\theta):={1\over (2M-\tilde{k}\sin \theta)},\quad p_r(\theta):=p_0(\theta){\mathrm{d}\over\mathrm{d}\theta}p_{r-1}(\theta), \quad r=1,2,\dots,
\end{align}
and $\left\vert R_J(M,K)\right\vert \leq C_J\tilde{k}M^{-2J-4},$ and $C_J$ is independent of $M$ and $\tilde{k}.$ If $\theta=0,$ the first four coefficients can be formulated as follows:
\begin{align}
p_0(0):={1\over 2M},\quad p_1(0):={\tilde{k}\over (2M)^3},\quad p_2(0):={3\tilde{k}^2\over (2M)^5},\quad p_2(0):={(15\tilde{k}^2-4M^2)\tilde{k}\over (2M)^7}.
\end{align}
We summarise the ideas above in Algorithm~\ref{algo:W2}.
\begin{algorithm}[h]
\caption{Algorithm: for $\tilde{k}<n<N$ (second phase)}\label{algo:W2}
\begin{algorithmic}[1]
\State Set $n_0=\ceil{\tilde{k}}$\;
\State Take $M\geq \max(n_0/2,N/2)$ sufficiently large and compute $\rho_{2M}(\tilde{k})$ using (\ref{eqn:asym_rho})\;
\State Construct $A_M(\tilde{k})$, $b_M(\tilde{k})$ as in and solve a linear system of equations:
 $$A_{M}(\tilde{k})\boldsymbol{\rho}_M(\tilde{k})= \mathbf{b}_M(\tilde{k})$$
 to obtain a vector of $\boldsymbol{\rho}_M(\tilde{k})$\;
 \State Set $w_n(\tilde{k}):=\gamma_n(\tilde{k})-{n\over ik}\rho_n(\tilde{k}),\quad n=n_0,\dots,N.$
 \end{algorithmic}
\end{algorithm}
\begin{remark}
Based on all the algorithms proposed by \citet{Dom_Gra:2011}, the FCC rule applied to solve (\ref{eqn:w_nFCC}) only requires $\mathcal{O}(N\log N)$ operations.
\end{remark}

%
%

\section{Table of cumulants}\label{sec:cums}
In Table~\ref{table:cumulants}, we show the first $c_1,$ second $c_2,$ and fourth $c_4$ cumulants of the GB model, the NIG model, the VG model and the $\rm CGMY$ model. In the $\rm CGMY$ model, we only present the cumulants when $Y\in(0,2)/\{1\}$ because when $Y=1,$ it becomes the VG model. Given the characteristic functions, the cumulants can be generally computed by using
$$c_k={1 \over i^k}{\partial^k \log \varphi(z)\over \partial z^n}\bigg\vert_{z=0}.$$
\begin{table}
\caption{The first $c_1,$ second $c_2,$ and fourth $c_4$ cumulants of various models.} \label{table:cumulants} 
\centering 
\small\addtolength{\tabcolsep}{-1pt}
\begin{tabular}{|l|l|} 
\hline
\multicolumn{2}{|l|}{L\'evy models}\\
\hline 
BS&$c_1=(r-q+\omega)t$\quad$c_2=\sigma^2 t,$\quad $c_4=0,$ $\omega=-0.5\sigma^2$\\
\hline
NIG&$c_1=(r-q+\omega)t+\delta t\beta/\sqrt{\alpha^2-\beta^2}$\\
\,&$c_2=\delta t \alpha^2 (\alpha^2-\beta^2)^{-3/2}$\\
\,&$c_4=\delta t \alpha^2 (\alpha^2+4\beta^2)^{-3/2}(\alpha^2-\beta^2)^{-7/2}$\\
\,&$\omega=-0.5\sigma^2-\delta(\sqrt{\alpha^2-\beta^2}-\sqrt{\alpha^2-(\beta+1)^2}) $\\
\hline 
VG&$c_1=(r-q+\theta+\omega)t$\\
\,&$c_2=(\sigma^2+\upsilon\theta^2)t$\\
\,&$c_4=3(\sigma^4\upsilon+2\theta^4\upsilon^3+4\sigma^2\theta^2\upsilon^2)t$\\
\,&$\omega={1 / \upsilon}\log(1-\theta\upsilon-\sigma^2\upsilon/2)$\\
\hline 
$\rm CGMY$&$c_1=(r-q+\omega)t$\\
\,&$c_2=(C\Gamma(2-Y)(M^{Y-2} + G^{Y-2})t$\\
\,&$c_4=(C\Gamma(4-Y)(M^{Y-4} + G^{Y-4})t$\\
\,&$\omega=\left(C\Gamma(-Y)G^Y\left(\left(1+\frac{1}{G}\right)^Y-1-\frac{Y}{G}\right)+C\Gamma(-Y)M^Y\left(\left(1-\frac{1}{M}\right)^Y-1+\frac{Y}{M}\right)\right)$\\
\hline 
\end{tabular}
\end{table}

\section*{Acknowledgement}
We thank Professor Bengt Fornberg, Department of Applied Mathematics, University of Colorado for teaching the singular Fourier--Pad\'e method and Victor Dominguez, Department of Mathematics, University of Navarra for help and advice on using the Filon--Clenshaw--Curtis rules.

\end{document}